\documentclass{aa} 
\usepackage{graphicx}
\usepackage{txfonts}
\usepackage{mwe,tikz}
\usepackage{hyperref}
\hypersetup{colorlinks=true,linkcolor=blue,citecolor=blue,filecolor=blue,urlcolor=blue}
\usepackage{tikz}
\usepackage{siunitx}
\usepackage{longtable}
\newcommand{\msunpyr}{\ifmmode{\,M_{\odot}\,\mbox{yr}^{-1}} \else{ M$_{\odot}$/yr}\fi}

\newcommand{\kms}{\ifmmode{\,\mbox{km}\,\mbox{s}^{-1}}\else{km\,s^{-1}}\fi}
\newcommand{\kpc}{\ifmmode {\,\mbox{kpc}} \else{kpc}\fi}
\newcommand{\msun}{\ifmmode M_{\odot} \else M$_{\odot}$\fi}
\newcommand{\rsun}{\ifmmode R_{\odot} \else R$_{\odot}$\fi}
\newcommand{\lsun}{\ifmmode L_{\odot} \else L$_{\odot}$\fi}
\newcommand{\zsun}{\ifmmode Z_{\odot} \else $Z_{\odot}$\fi}
\newcommand{\xsun}{\ifmmode X_{\odot} \else $X_{\odot}$\fi}
\newcommand{\velo}{\ifmmode\varv\else$\varv$\fi}
\newcommand{\vinf}{\ifmmode\velo_\infty\else$\velo_\infty$\fi}
\newcommand{\rgal}{\ifmmode \,R_{\mathrm{gal}} \else R$_{\mathrm{gal}}$\fi}

\begin{document} 

    \title{First measurement of wind line formation regions in an early O-type star}

   \subtitle{The eclipsing binary system AzV 75 in the Small Magellanic Cloud}

   \author{
          D. Pauli \inst{1}\fnmsep\thanks{daniel.pauli@kuleuven.be}$^{,\dagger}$ \and T.N. Parsons
          \inst{2}\fnmsep\thanks{timothy.parsons.15@ucl.ac.uk}$^{,\dagger}$           \and R.K. Prinja\inst{2}
          }
             \institute{Institute of Astronomy, KU Leuven, Celestijnenlaan 200D, 3001 Leuven, Belgium\\
        \and
   University College London, Department of Physics and Astronomy, Gower Street, London WC1E 6BT, UK\\ \\
              $^\dagger$ Equal contribution is indicated by shared first-authorship. \\
             }

   \date{Received 24 April 2026; accepted 16 Juli 2026}
 
   \abstract
   {Massive stars with their strong ionizing radiation and strong stellar winds are the key feedback agents of the Universe. Stellar winds of massive stars are often measured by fitting resonance lines in the UV using non-local thermal equilibrium (non-LTE) stellar atmosphere models.}
   {We aim to conduct the first measurement of the resonance line formation regions in an early-type eclipsing binary in the Small Magellanic Cloud, namely AzV 75.}
   {We employed photometric data from TESS and ASAS-SN in combination with radial velocity measurements from multi-epoch Hubble Space Telescope UV spectra to derive the ephemeris. We examined the intensity changes in the \ion{C}{IV} and \ion{N}{V} resonance lines in the UV and combined them with a light-curve analysis to estimate the region in the wind where these lines are formed.}
   {We derived that AzV 75 has an orbital period of $P_\mathrm{orb}=\SI{165.925 \pm0.115}{d}$, eccentricity of $e=0.42\pm0.02$, mass ratio of $q = 0.72\pm0.03$, and inclination of $i=85.77^\circ$. With this orbital configuration, no secondary eclipse is expected. We report that the optically thick UV resonance lines exhibit flattening and shortening of the absorption trough, and weakening of their emission features, as they approach the phase of the expected secondary eclipse, while the continuum UV flux appears to remain unaffected. We illustrate that this behaviour can be well explained by the primary's optically thick wind eclipsing the secondary star. The \ion{C}{IV} and \ion{N}{V} resonance line formation regions in the primary star extend up to $316\,\rsun$.}
   {We have measured the line formation regions of resonance lines in a stellar wind of an early O-type star. These are important benchmarks for 1D as well as 3D non-LTE stellar atmosphere models. Even though we have numerous multi-epoch UV spectra at hand, the eclipse is not fully sampled, leaving some freedom in the derived sizes of the line formation regions. A first comparison to 1D stellar atmosphere models indicates that a classical $\beta$ law with an exponent of $\beta=0.5$ instead of $\beta=0.8$ might be favoured for the primary star's velocity field.
   }


    \titlerunning{First measurement of wind line formation regions in an early O-type star}
    \maketitle
%

\section{Introduction}

Massive stars ($M_\mathrm{ini}\gtrsim 8 M_\odot$) emit copious amounts of ionizing radiation and power strong UV-radiation driven stellar winds, making them the principal drivers of enrichment and evolution of the interstellar medium and of galaxies more generally \citep[e.g.][]{Barkana2006, Hopkins2014}. Despite their importance, stellar winds of hot massive stars are not well understood. Often, theoretical predictions and empirical measurements of the mass-loss via stellar winds diverge \citep[e.g.][and references therein]{Pauli2025}. Often, the mass-loss rate is measured by fitting UV resonance lines with non-local thermal equilibrium stellar atmosphere models. While the characteristic P\,Cygni profiles themselves provide some information on the properties of the wind, such as the velocity or density structure \citep{Kudritzki2000}, direct measurements of the line formation regions of these resonance lines in the outer stellar atmosphere are still missing. 

Frequently, massive stars are seen to have at least one close stellar companion. Earlier studies have shown that, in the Milky Way at least, the great majority of massive stars exist in binary or higher-order multiple systems \citep[][among others]{Sana2012, Chini2012, Kobulnicky2014, Marchant2024}. The predominance of binarity among massive stars has been shown to extend to lower metallicities in nearby dwarf galaxies, in particular, the Small Magellanic Cloud (SMC) \citep[e.g.][]{Bodensteiner2025, Bestenlehner2025b,Sana2025}. 

Eclipsing binaries are a very powerful tool in stellar astrophysics, as they provide accurate measurements of the mass of the two stars as well as an independent measurement of the stars' radii and temperature ratio. In a long-period binary, which is wide enough such that the stars have not interacted yet, and which contains a hot star with a (partially) optically thick wind and a UV bright companion with a weak or absent wind, the wind of the strong-winded star should eclipse the companion. Measuring this eclipse duration will enable the size of the wind itself to be measured. However, the chance of having an eclipse decreases with increasing period, drastically reducing the available sample of target stars for this purpose. Furthermore, one needs a star with a strong wind dominating in the UV but a companion bright enough to see changes in the UV resonance lines when the companion is occulted. This is more likely to be fulfilled at lower metallicity, where stellar winds are weaker. Lastly, one needs multi-epoch observations sampled roughly evenly over the orbit to see the phases at which the resonance lines block the flux from the companion.

The SMC massive star AzV 75, previously described as spectral type O5 III(f+) \citep{Walborn2000} but recently revised as O3.5 III(f) by \citet{Bestenlehner2025a}, is an evolved massive star that forms part of the extensive database of low-metallicity massive stars observed by the \textit{Hubble Space Telescope} (HST) and assembled under the HST legacy programme: the Ultraviolet Legacy Library of Young Stars as Essential Standards (ULLYSES) \citep{Roman2025}. The massive star portion of that programme has provided an extensive resource of medium-high resolution UV spectra, principally obtained using the STIS and COS instruments aboard HST. Along with other spectral data, the libraries of data produced are essential materials for the determination of many physical parameters and characteristics of massive stars at low and very low metallicities \citep[see, in particular, the \textit{XShootU} programme and spectral database described in][]{Vink2023, Sana2024}.

Just recently, AzV 75 was indicated to be an eclipsing binary system by a statement of \citet{Pedersen2025}, based upon evidence in \textit{NASA Transiting Exoplanet Survey Satellite} \citep[TESS;][]{Ricker2015} data. Detailed evidence of binarity, as shown by radial velocity (RV) shifts of two stellar components and variation in the wind lines in individual UV spectra, some of which form the basis of the ULLYSES High Level Science Product (HLSP) UV spectrum of AzV 75, was first presented by \citet{Parsons2026}. 
Furthermore, multiple UV spectra of AzV 75 have been obtained over a period of several years as part of HST instrument calibration exercises under a number of different programmes, which are ongoing. The most recent spectrum, up to the present date, was obtained on 2025 October 24 (2460968 JD). 
All of this makes AzV 75 an ideal candidate to measure for the first time the size of the line formation regions of the optically thick resonance lines in the UV.

\section{Analysis of the binary orbit}
\subsection{Method}
\subsubsection{Light curve}
    \label{sec:LC}

    \citet{Pedersen2025} mentioned that AzV 75 exhibits an eclipse in the TESS data, but no period was determined. We reexamined the TESS data of AzV 75, which was observed during sectors 1, 27, 28, 67, 68, 94, and 95. Sector 28 shows an apparent single instance of a primary eclipse being recorded at a heliocentric Julian date (HJD) of 2459072.3, which also reappears in sector 95 at 2460897.6\,HJD. The duration of the eclipse is approximately 1.3 days. TESS has a low spatial resolution of $\SI{21}{\arcsec\,px^{-1}}$ and in sector 28 the pixel in which AzV 75 appears contains at the edge one other source of comparable brightness (see Fig. \ref{fig:sector28}). Therefore, we refrained from using the photometric data from sector 28 for a later light-curve analysis. In sector 95, AzV\,75 is the only bright source in its corresponding pixel (see Fig.~\ref{fig:sector95}), allowing the extracted light curve to be used during the light curve analysis.

\begin{figure}
    \centering
        \includegraphics[trim= 0.5cm 0cm 1.5cm 1cm ,clip,width=\linewidth]{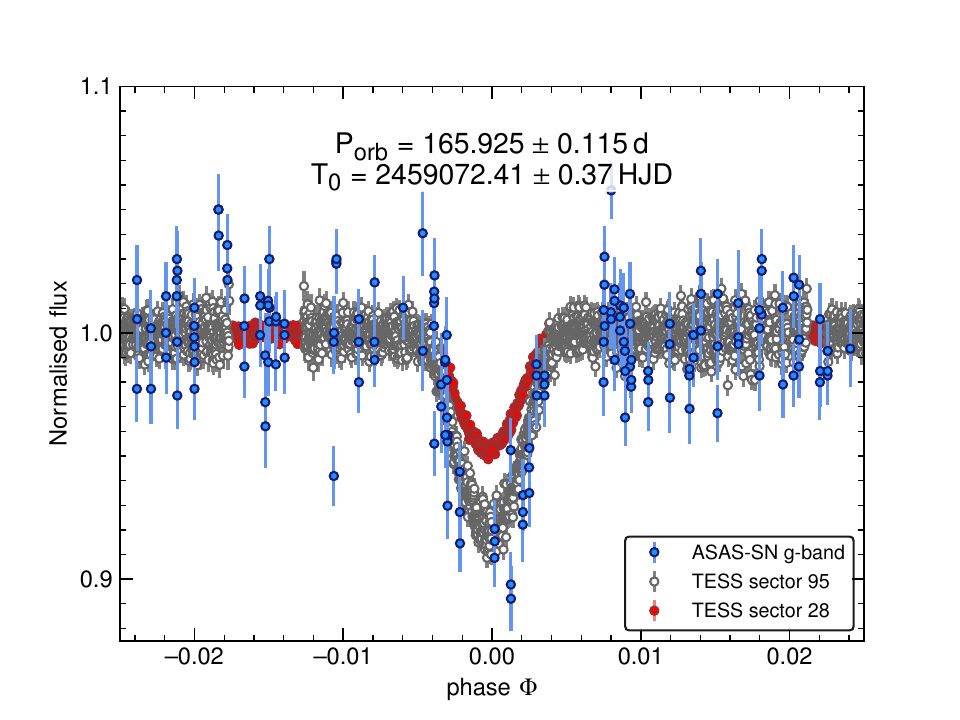}
        \caption{Phased TESS and ASAS-SN photometric data close to zero phase. Note that during TESS sector 28 the pixel in which AzV~75 appears contains another bright star, leading to an apparent weaker eclipse.}
        \label{fig:LCzoom}
    \end{figure}

    To obtain an orbital period of the system, we extracted All-Sky Automated Survey for Supernovae (ASAS-SN) \textit{g}-Band photometric aperture data \citep{Shappee2014,Kochanek2017,Hart2023}. The ASAS-SN survey has a cadency of 2 to 3 days and has photometric coverage between 2458850.61 and 2460848.68 HJD. From a visual inspection, we identified four primary eclipses occurring at 2459072.4$\pm1$ HJD, 2459238.5$\pm1$ HJD, 2459570.1$\pm1$ HJD, and 2459902.3$\pm1$ HJD, including the eclipse seen in sector 28 of TESS. The identified primary eclipses in the TESS and ASAS-SN data indicate a period of $165.925\pm0.115$ days. The high-quality TESS data contains sectors at about half-phase. No eclipse from either the primary or the secondary can be detected here. From the TESS and ASAS-SN observations, we obtain a reference eclipse timing of $T_{0} = 2459072.41 \pm0.37$ HJD. The phased primary eclipse is shown in Fig.~\ref{fig:LCzoom}.
    
\subsubsection{RVs}
\label{sec:RV}

    The HST spectra of AzV~75 show a clear line splitting in the \ion{O}{IV}\,$\lambda\lambda1338.61,1342.99,1343.51$ photospheric lines (see figure~5 in \citealt{Parsons2026} and, here, Fig.~\ref{fig:phases}). While most of the HST UV spectra were already wavelength-calibrated, for a small number of spectra, calibration was required. Hence, to obtain the RVs from the HST data, we first cross-correlated all spectra using the \ion{C}{II} 1335 \AA{} interstellar line to derive a correction factor, aligning the spectra in the same reference frame. As a mask spectrum, we used the HST STIS spectrum taken at 2451322.267\,HJD. The applied corrections are listed in Table~\ref{tab:RVs}. To measure the RV shifts of each star, we employed a Markov chain Monte Carlo (MCMC) approach. Within this method, we shifted two individual line functions of the primary and secondary -- obtained by fitting random synthetic stellar spectra to the \ion{O}{IV} multiplet at maximum separation -- by a given set of RVs. The combined spectrum was then compared to the observed one, and the fit's quality was evaluated using a least-squares likelihood function. The MCMC can quickly explore the full parameter space and converges after a few hundred steps on the true solution. We sampled around the true solution for 1000 RV combinations, which provided us with the posterior distribution. The quoted error margins correspond to the 68\% confidence interval of the posterior. A complete list of the derived RVs for each spectrum is provided in Table~\ref{tab:RVs}, and the fits of the individual spectra can be inspected in Fig.~\ref{fig:phases}.
    
    AzV~75 was observed three times in the optical with X-Shooter \citep{vernet2011}. Unfortunately, all of these observations have been taken close to an eclipse, when the RVs are small. Therefore, we are unable to see in these spectra any line splitting or deformation, and thus to derive reliable RVs.

\subsection{Modelling the light and RV curve}
\label{sec:PHOEBE}
    
    To model the light and RV curve, we employed the Physics of Eclipsing Binaries (PHOEBE) code \citep{Prsa2016,Conroy2020}, version 2.4. 
    Within the photometric data, we do not see a secondary eclipse occurring. This means that the light curve cannot be used to obtain the eccentricity and the argument of the periastron passage, and that these parameters can only be obtained from the fit of the RV curve. Therefore, we split the fitting procedure into two steps.

\begin{figure}
    \centering
    \includegraphics[trim= 0.9cm 1.3cm 1.7cm 2.5cm ,clip,width=\linewidth]{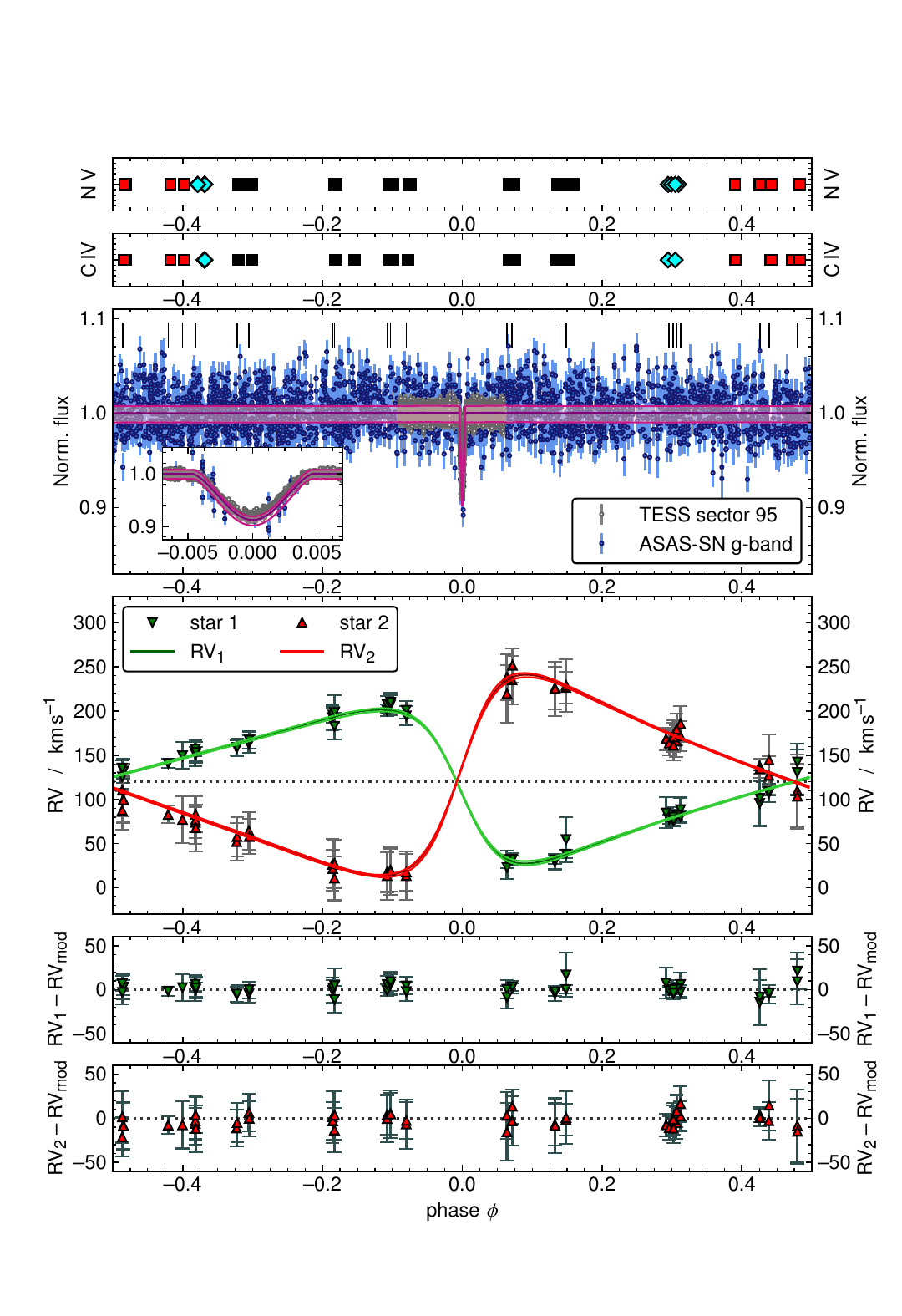}
    \caption{AzV 75 observed light and RV curve compared to the best-fitting PHOEBE model. \textit{Upper panels:} Variability of the \ion{N}{V} and \ion{C}{IV} resonance lines in the HST UV spectra. Phases at which HST spectra show a weak, intermediate, and strong emission feature in a resonance line of interest are indicated by red, cyan, and black symbols, respectively (see also Fig.~\ref{fig:PCygni}). \textit{Middle panels:} Light curves derived from ASAS-SN \textit{g}-band observations (blue) and TESS sector 95 (grey) compared to the best-fitting PHOEBE model and its uncertainties (pink lines and shaded region). The inset shows a zoom-in on the primary's eclipse, which is fully resolved in the TESS data. Phases of the HST UV spectra used to derive the RVs are indicated by vertical lines. Measured RVs of the primary (upside-down green triangles) and secondary (red triangles) are overplotted by the PHOEBE model and its uncertainties (lines and shaded regions). The dashed line indicates the systemic velocity of the system. \textit{Lower panels:} Residuals of the RV curve for each star.}
    \label{fig:SpecOrbit}%
    \end{figure}

    At first, the RV curve was fitted using the built-in MCMC sampler option of the PHOEBE code. For the fitting procedure, the orbital period, $P_\mathrm{orb}$, and time of the primary eclipse, $T_0$, were fixed at the values derived from the TESS and ASAS-SN light curves (see Sect.~\ref{sec:LC}). The free parameters are the projected semi-major axis, $a\sin i$, eccentricity, $e$, argument of the periastron, $\omega$, the systemic velocity, $\varv_\gamma$, and the mass ratio, $q=M_2/M_1$. A flat prior in the mass ratio of $q=0.4\,\textit{--}\,1.5$ and in the systemic velocity of $\varv_\gamma=\SIrange{100}{200}{km\,s^{-1}}$, allowing for variation around the systemic velocity of the SMC, were employed to help with convergence. The fit of the RV curve is shown in Fig.~\ref{fig:SpecOrbit}, and the derived orbital parameters are presented in Table~\ref{tab:PHEOEB_orbit}. The posterior distributions are displayed in Fig.~\ref{fig:RVposteriors}.

    \begin{table}[tbp]
 		\centering
 		\caption{Orbital parameters obtained from the light and RV curve.}
 		\label{tab:PHEOEB_orbit}
 		\begin{tabular}{lcc}
 			\hline\hline \rule{0cm}{2.8ex}
 			\rule{0cm}{2.8ex}Parameter & \multicolumn{2}{c}{Value}\\ 
 			\hline\rule{0cm}{2.8ex}
 			$P\,(\mathrm{d})$ & \multicolumn{2}{c}{$165.925\,\,\,\,(\mathrm{fixed})$}\\ 
 			\rule{0cm}{2.8ex}HJD$_0$ & \multicolumn{2}{c}{$2459072.41\,\,\,\,(\mathrm{fixed})$}\\ 
 			\rule{0cm}{2.8ex}$q$ & \multicolumn{2}{c}{$0.77^{+0.04}_{-0.04}$}\\ 
 			\rule{0cm}{2.8ex}$e$ & \multicolumn{2}{c}{$0.48^{+0.02}_{-0.02}$}\\ 
 			\rule{0cm}{2.8ex}$\omega\,(\mathrm{^\circ})$ & \multicolumn{2}{c}{$97.4^{+1.8}_{-1.7}$}\\ 
 			\rule{0cm}{2.8ex}$a\sin i\,({R_\odot})$ & \multicolumn{2}{c}{$579^{+15}_{-16}$}\\ 
 			\rule{0cm}{2.8ex}$i\,(\mathrm{^\circ})$ & \multicolumn{2}{c}{$85.77^{+0.20}_{-0.49}$}\\ 
 			\rule{0cm}{2.8ex}$a\,({R_\odot})$ & \multicolumn{2}{c}{$581^{+15}_{-16}$}\\ 
            \rule{0cm}{2.8ex}$\varv_\gamma\,(\mathrm{km\,s^{-1}})$ & \multicolumn{2}{c}{$120.1^{+1.2}_{-1.2}$}\vspace{0.15cm}\\
 			\hline
 		\end{tabular}
 		\vspace*{3ex}
 		\caption{Stellar parameters obtained from the light and RV curve.}
 		\label{tab:PHOEBE_stellar}
 		\begin{tabular}{lcc}
 			\hline\hline \rule{0cm}{2.8ex}
 			\rule{0cm}{2.8ex}Parameter & Primary & Secondary\\ 
 			\hline
 			\rule{0cm}{2.8ex}$T_\mathrm{eff}\,\,[\mathrm{kK}]$ & $36\,\,(\mathrm{input})$ & $46\,\,(\mathrm{input})$\\ 
 			\rule{0cm}{2.8ex}$M\sin^3i\,\,[{\msun}]$ & $53.6^{+4.4}_{-4.5}$ & $41.1^{+3.5}_{-3.6}$\\ 
 			\rule{0cm}{2.8ex}$M\,\,[{\msun}]$ & $54.0^{+4.5}_{-4.5}$ & $41.5^{+3.5}_{-3.6}$\\ 
 			\rule{0cm}{2.8ex}$R\,\,[{\rsun}]$ & $25.45^{+0.25}_{-0.82}$ & $9.65^{+2.63}_{-0.71}$\\ 
 			\rule{0cm}{2.8ex}$\log g\,\,[\si{cm\,s^{-2}}]$ & $3.36^{+0.04}_{-0.04}$ & $4.09^{+0.22}_{-0.08}$\vspace{0.15cm}\\ 
 			\hline 
 		\end{tabular}
 	\end{table} 
        
    From the fitting of stellar eclipses, one can, in principle, obtain information about the stellar temperatures as well as the stellar radii. However, the ASAS-SN and the TESS photometric data are taken in almost the same pass-band and only show primary eclipses, meaning that we cannot obtain reliable temperatures of both stars. With AzV~75 containing one of the brightest stars in the SMC and being unrecognized as a binary system, it has been subject to multiple studies obtaining the stellar parameters as if it were a single star \citep{Bouret2021,Bestenlehner2025b}. However, the binary companion, which contributes about 25\% to the UV flux (see Sect.~\ref{sec:wind_variability}), is expected to also have a noticeable signature in the optical, questioning the reliability of their temperature estimate for the primary star of this system. Unfortunately, all the available optical spectra were taken by chance close to an eclipse, preventing us from seeing line deformations, and thus from performing a detailed stellar atmosphere fitting, providing accurate stellar temperatures. 
   
    To get an estimate of the stellar temperature of the stars, we used the information from the RV fit in combination with the spectral energy distribution (SED) of the system to find two approximately representative stellar atmosphere models. To achieve this,  the publicly available Potsdam Wolf-Rayet stellar (PoWR) OB-Vd3 stellar atmosphere model grids of \citep{Pauli2025b} were employed. These grids are based on the MIST \citep{Choi2016} single star evolution models and use the \citet{Vink2001} mass-loss rates divided by a factor of 3. From these models, we can roughly translate the orbital masses of the stars, the total luminosity of the system, and the contribution of the two stars to the UV flux to their surface temperatures. 
    
    The closest PoWR models at SMC metallicity that reproduce the aforementioned parameters are the `36-34' model of the SMC-OB-Vd3 grid for the primary and `46-42' of the SMC-OB-Vd3 grid for the secondary. The primary's model has a temperature of $T_{\ast,\,1}=\SI{36}{kK}$, a surface gravity of $\log(g_{1}/(\mathrm{cm\,s^{-2}}))=3.4$, a luminosity of $\log(L_1/L_\odot)=6.02$, and a mass-loss rate of $\log(\dot{M_1}/(\msunpyr))=-6.30$. The model temperature of the secondary is $T_{\ast,\,2}=\SI{46}{kK}$, with surface gravity of $\log(g_2/(\mathrm{cm\,s^{-2}}))=4.2$, luminosity $\log(L_2/\lsun)=5.43$, and mass-loss rate $\log(\dot{M_2}/(\msunpyr))=-7.2$. In Fig.~\ref{fig:PoWR}, the individual and combined synthetic model spectra are plotted against the system's SED and a UV spectrum, illustrating that the chosen models are representative of the stars in AzV~75.
    
    For the fit of the light curve, we also employed the MCMC sampler of PHOEBE. Again, the orbital period, $P_\mathrm{orb}$, and time of the primary eclipse, $T_0$, were fixed to the values derived from the TESS and ASAS-SN light curves. Unfortunately, PHOEBE does not contain synthetic spectra for hot massive stars yet, so we had to approximate the spectra of the stars as blackbodies. To describe the effect of limb darkening, we extracted the emergent flux distribution from the PoWR models in the g band and fitted them using a quadratic limb-darkening law \citep{DiazCordoves1992}, yielding coefficients of $a_1=0.00583$, $b_1=-0.00332$, and $a_2=0.17$, $b_2=-0.09$ for the primary and secondary, respectively. For the fitting procedure, we employed the ASAS-SN g-band light curve, as well as the TESS light curve extracted from sector 95. Since the TESS photometry is not flux calibrated, we normalized the light curves and used the `dataset-coupled' option of PHOEBE. The free fitting parameters are the inclination, $i$, of the system and the stellar radii of the primary, $R_1$, and the secondary, $R_2$. The resulting fit of the light curve is shown in the top panel of Fig.~\ref{fig:SpecOrbit} and the resulting parameters are listed in Table~\ref{tab:PHOEBE_stellar}. The posterior distributions of the MCMC sampler are displayed in Fig.~\ref{fig:LCposteriors}.

\section{Wind variability}
\label{sec:wind_variability}

    \subsection{Stellar wind line intensity variations}
    \label{sec:wind_variability_sub}
    
   \begin{figure}
   \centering  
        \begin{tikzpicture}
            \node[anchor=south] at (0, 6) {\includegraphics[trim= 1.cm 0.2cm 2cm 1.3cm ,clip,width=\linewidth]{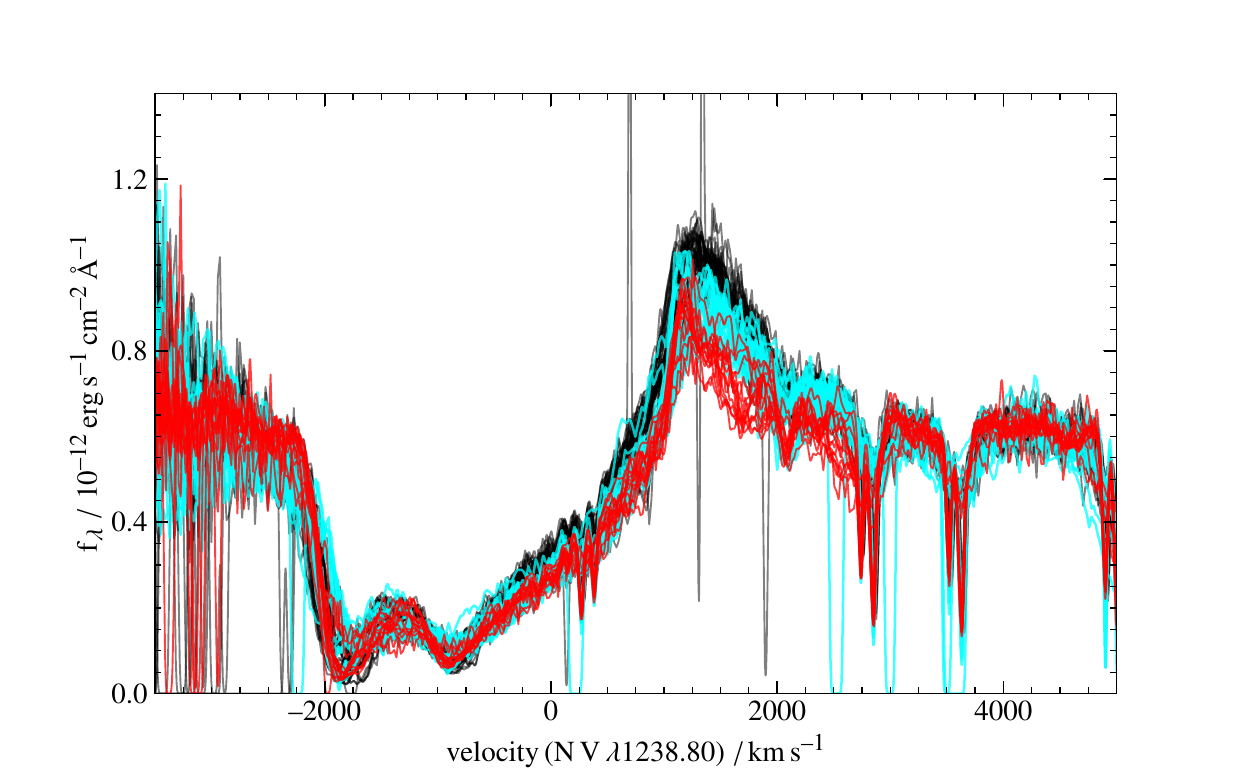}};
            \node[fill=white, inner sep=2pt] at (0, 11.4) {\ion{N}{V} -- 1238.80 (1242.80)\,\AA}; 
            \node[anchor=south] at (0, 0) {\includegraphics[trim= 1.cm 0.2cm 2cm 1.3cm ,clip,width=\linewidth]{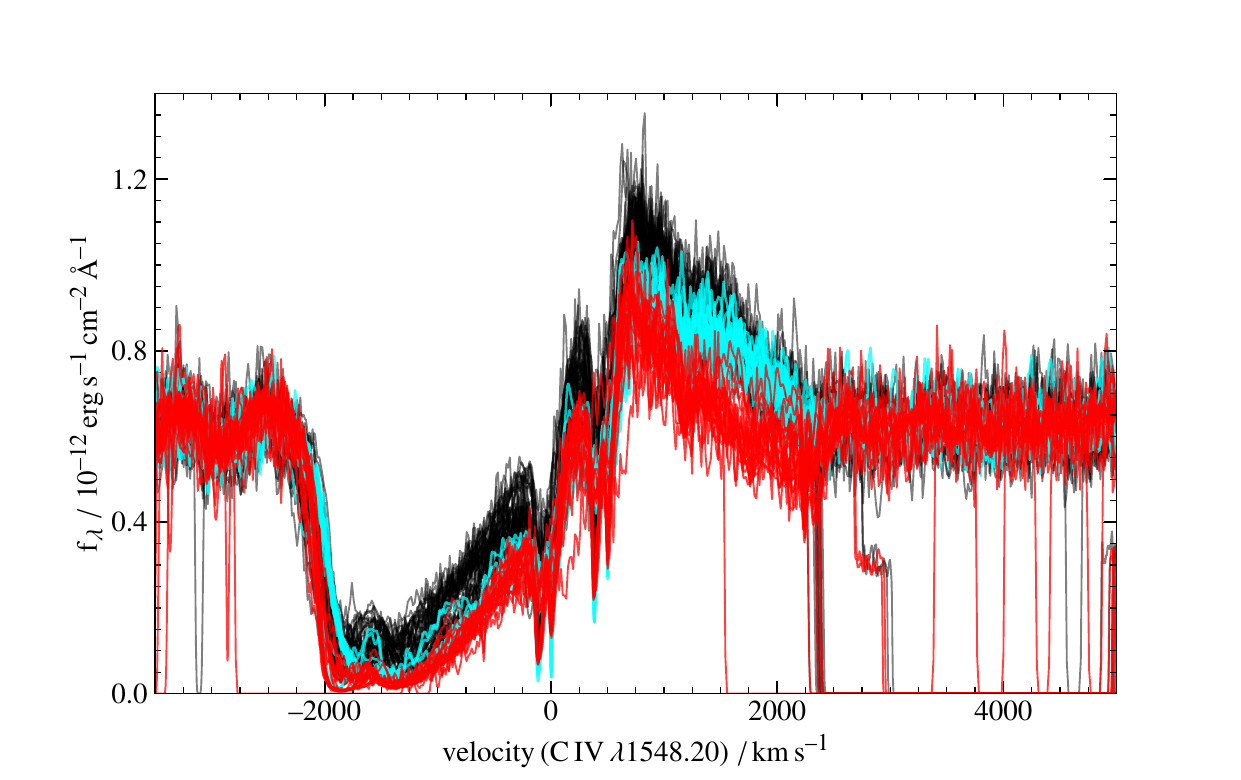}};
            \node[fill=white, inner sep=2pt] at (0, 5.4) {\ion{C}{IV} -- 1548.20 (1550.78)\,\AA}; 
        \end{tikzpicture}
        \caption{Detail of the wind-formed \ion{N}{V} (upper panel) and \ion{C}{IV} (lower panel) resonance line P Cygni-type profiles taken from the available HST UV spectra. All spectra are shifted into the rest frame of the primary and are corrected for the systemic velocity. The \ion{N}{V} is corrected for the interstellar Ly$\alpha$ blend by modelling a (fully damped) profile of column density of $\SI{6.9e21}{cm^{-2}}$.  
        Note that the colour coding matches the phase-based colouring of spectra in the upper panels of Fig.~\ref{fig:SpecOrbit}. }
         \label{fig:PCygni}
   \end{figure}

    In addition to the RV shifts, the HST spectra show varying stellar wind-driven emission features in the P\,Cygni profiles of the \ion{N}{V} and \ion{C}{IV} resonance line profiles \citep{Parsons2026}. This occurs consistently across both species. In Fig. \ref{fig:PCygni} all available HST UV spectra containing the resonance lines are displayed. One can see that the spectra show three characteristic states: i) maximum emission, ii) a $\approx20\%$ drop in flux in the emission feature, and iii) intermediate stages with reduced emission. For the \ion{C}{IV} resonance line, one can additionally see that at the different states, the change in emission occurs simultaneously with, respectively, i) an in-filled absorption trough, ii) a fully saturated absorption trough, and iii) a weaker but still infilled absorption trough. The \ion{N}{V} resonance line (corrected for interstellar Ly$\alpha$ absorption) exhibits a similar behaviour in its emission component. However, its absorption trough, which is not fully saturated, and thus indicative of a somewhat lower optical depth compared to \ion{C}{IV}, shows less pronounced variability. 
    This reduced variability can be attributed to the reduced received flux caused by the blend with the interstellar Ly$\alpha$ line, leading to a lower signal-to-noise ratio in the \ion{N}{V} absorption trough. As already noted by \cite{Patel2024}, in contrast to the wind lines, the UV continuum of all spectra remains unchanged (noting that there is no UV spectrum obtained at a time coincident with a primary eclipse).

\begin{figure}
        \centering  
        \includegraphics[trim= 1.cm 0.2cm 2cm 1.2cm ,clip,width=\linewidth]{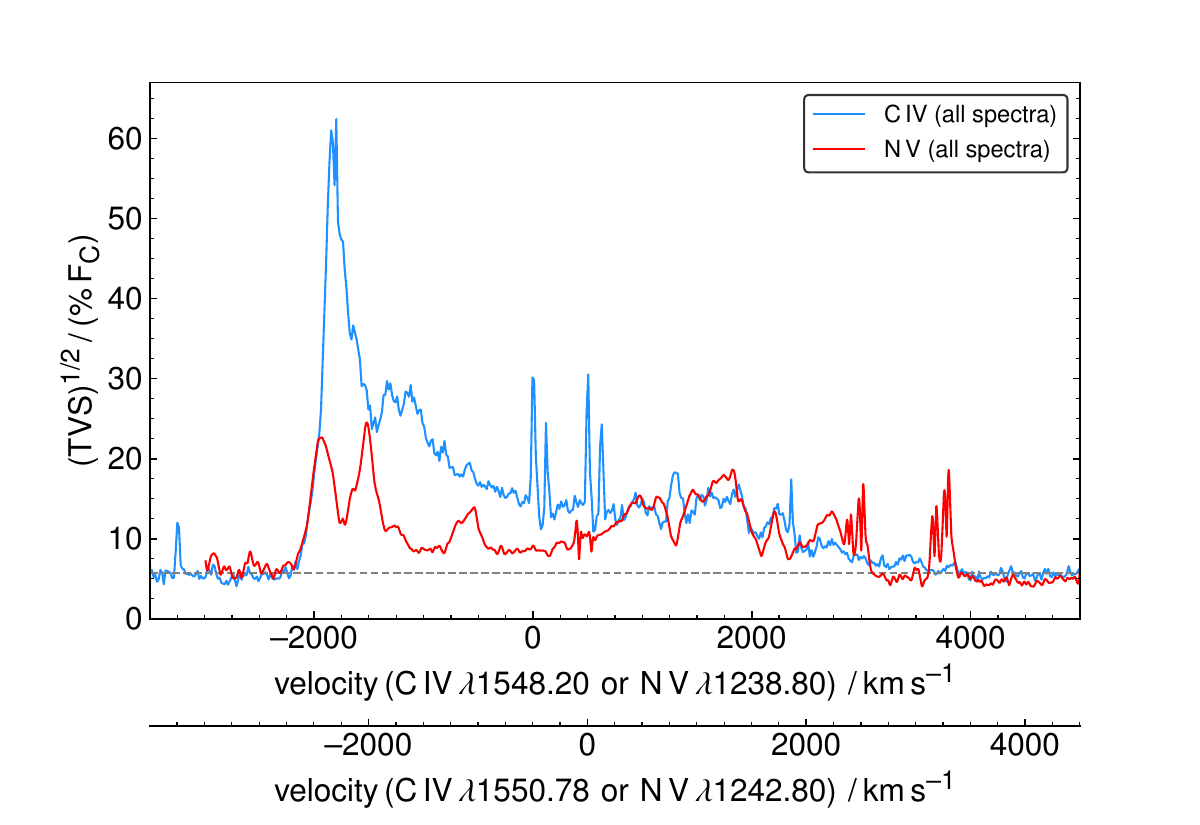}
        \caption{TVS of the \ion{N}{V} (red) and \ion{C}{IV} (blue) resonance lines in velocity space of the \ion{N}{V} $\lambda1238.80$ ($\lambda1242.80$, lower ticks) or \ion{C}{IV} $\lambda1548.20$ ($\lambda1550.70$, lower ticks) lines, respectively. In both cases, the variance rapidly decreases towards redward velocities, while in the blueward absorption trough one can see some variance across the spectra, typical for a O-type star. The overall similarity in both the shape of the variance and the velocity range over which it is observed strongly suggests that the two resonance lines are formed in the same regions of the primary’s wind.}
         \label{fig:TVS_NV}
    \end{figure}

    Both, the \ion{N}{V} and the \ion{C}{IV} resonance line have a terminal wind velocity around $v_\infty\approx\SI{1960\pm20}{km\,s^{-1}}$. This value is slightly smaller than earlier measurements \citep[e.g.][]{Bouret2021,Hawcroft2024b,Parsons2026}, because those did not account for the previously unknown orbital motion of the primary star. The fact that both resonance lines exhibit absorption troughs developed to the same velocity is already a first indication that the lines are formed in a similar region of the primary's wind. To further investigate this, we computed temporal variance spectra \citep[TVS;][]{Fullerton1996} that quantify the variance across all spectra in contrast to the mean spectrum, allowing us to search for additional variability in stellar wind lines. The resulting TVS for the Ly$\alpha$-corrected \ion{N}{V} and \ion{C}{IV} resonance lines are depicted in Fig.~\ref{fig:TVS_NV}. In the emission component, the variability pattern of the two lines are nearly identical. In the absorption through the variability of the \ion{N}{V} is reduced, which we attribute to the weaker signal caused by the blend with the interstellar Ly$\alpha$ line. Nevertheless, the variability in this region still remains very similar to that from the \ion{C}{IV}, reinforcing the conclusion that both resonance lines are formed in comparable regions of the primary's wind.

    In other massive binaries in the SMC containing one strong-winded star with a weak-winded but still hot companion (such as AzV~476 \citep{Pauli2022b} or 2dFS~163 and 2553 \citep{Ramachandran2024}), one can often see that the secondary adds net UV flux to the \ion{C}{IV} line, leading to an infilling of the absorption trough, similar to what is seen in the black spectra of Fig.~\ref{fig:PCygni}. However, a simultaneous reduction of the emission peak and strengthening of the absorption trough for such a binary does not appear to have been previously reported. Intuitively, one might expect a reduction to occur during the secondary eclipse, but this scenario should also produce a similar decrease in the continuum flux, an effect that is not seen in the HST spectra. 

    The top panels of Fig.~\ref{fig:SpecOrbit} indicate which of the three characteristic \ion{N}{V} and \ion{C}{IV} profile states is observed at each orbital phase. A clear correlation emerges: the reduction in the intensity of the \ion{C}{IV} resonance line occurs only around orbital phases where the secondary eclipse should take place, but this reduction persists over a significantly longer fraction of the orbital cycle than a `normal' eclipse. There is also a lead-in and run-out of this feature at intermediate orbital phases.

    Given the strong phase dependence, we interpret the observed behaviour of the UV resonance lines as evidence for a wind eclipse. At quadrature, the UV flux of both stars is received, such that the secondary's UV continuum fills in the primary's otherwise fully saturated \ion{C}{IV} P\,Cygni profile, and also enhances the emission feature, as seen in the black spectra shown in Fig.~\ref{fig:PCygni}. Still, even when the secondary is projected in front of the primary's wind, one would not expect a significant reduction in the flux due to the large difference in the size of the secondary's star disc and the primary's wind ($R_2\ll R_\mathrm{1,\,wind}$). However, while most strong-winded O stars are optically thin in the continuum, they remain optically thick in their resonance lines. Consequently, once the secondary passes behind the primary’s wind, its contribution to the resonance-line flux becomes blocked, while the remaining UV continuum passes largely unattenuated through the optically thin wind, leading to a similar effect as seen in the red spectra presented in Fig.~\ref{fig:PCygni}. Furthermore, when the star is behind regions where \ion{C}{IV} is no longer completely optically thick, one expects intermediate stages such as during an ingress and egress in a normal eclipsing binary. These ingress and egress phases provide valuable information about the density structure of the star and future tailored observations might provide novel insights into the stellar wind physics.

    \subsection{Estimating the size of the \ion{C}{IV} and \ion{N}{V} resonance line formation regions}

    AzV~75 is a standard calibration star for HST COS \citep[starting with][]{HST_prop_AzV75}, meaning that it is observed regularly and that in the archives several high-quality UV spectra are available that, coincidentally, cover almost the full orbit. With the extracted information indicating the orbital phases at which the secondary is behind the optically thick regions of the resonance lines of the primary's stellar wind, it is possible to calculate the radial extent of these regions. This information can then be compared to stellar atmosphere models and provides a unique benchmark for stellar wind theory.

    Unfortunately, there is no eclipsing binary model tool available that accounts in detail for each star's stellar atmosphere and the connected changes in optical depth across the stellar wind. In the literature, several studies have addressed eclipses in Wolf–Rayet  binaries, where the strong winds of both components interact, often producing colliding‑wind X‑ray shocks and phase‑dependent line‑profile variability \citep[e.g.][]{Auer1994,Lamontagne1996,Perrier2009}. However, our system is quite different, as only the primary drives a strong wind, whereas the secondary does not appear to possess a significant outflow (see Sect.~\ref{sec:secondary_wind}).
    Under the assumption that the \ion{C}{IV} and \ion{N}{V} resonance line formation region in the primary's wind and the secondary can be approximated by uniformly radiating stellar spheres, and given that the radius difference between the primary's wind and the stellar disk of the secondary is much larger than a factor of ten, the duration of the eclipse can be approximated analytically.

    To calculate the duration of the total eclipse (i.e. when the secondary is completely behind the primary's wind), we calculated for each orbital phase the projected distance, $\rho(t)$, between the two binary components in an eccentric binary following the formula from \citet{Lewis2013}: 
    \begin{equation}
        \rho(t) = r(t)\sqrt{1-\sin(i)\sin(\omega+\theta(t))}~,
    \end{equation}
    where $r(t)$ is the radial distance between the two stars, $i$ is the system's inclination, $\omega$ the argument of periastron, and $\theta(t)$ the true anomaly. As soon as $\rho(t)$ becomes smaller than $R_\mathrm{1,\,wind}-R_2$, the secondary is fully covered by, or in front of, the primary's wind. 

   \begin{figure}
   \centering
   \includegraphics[trim= 0.9cm 0.9cm 1.7cm 1.9cm ,clip,width=\linewidth]{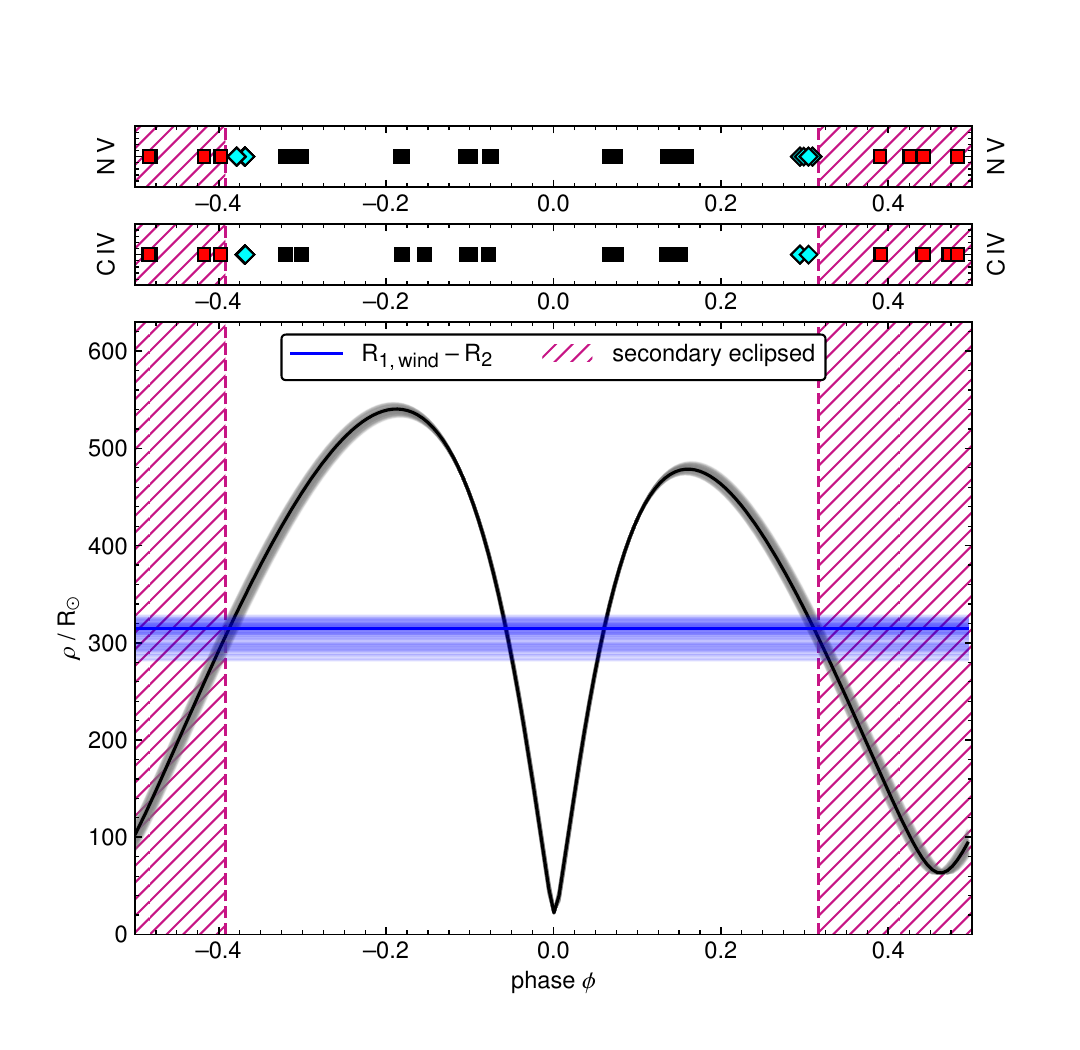}
      \caption{Phased projected distance between the primary's \ion{N}{V} and \ion{C}{IV} resonance line formation region and the secondary (black line). The uncertainties of the projected distance are illustrated as shaded grey regions around the best fit. The threshold distance, after which the secondary is fully in front of or behind the primary's wind, is indicated by a blue line, and the uncertainties are marked by the shaded blue region. The phase when the secondary is eclipsed by the primary's wind is indicated by the hatched pink area. Phases at which the HST UV spectra show a fully saturated, partially filled, and in-filled absorption trough of the \ion{C}{IV} resonance line are indicated by red squares, cyan diamonds, and black squares at the top of the figure, respectively. Note that only the red squares are associated with a total eclipse of the secondary by the primary's wind.
              }
         \label{fig:CIV_LC}
   \end{figure}

    From the orbital analysis, we already have well-constrained ephemerides, with the only exception being $\omega$, which has a bimodal posterior distribution. To fit the line-forming region of the resonance lines, we used an MCMC approach. We adopted from the orbital solution the semi-major axis, $a$, the eccentricity, $e$, the inclination, $i$, the orbital period, $P$, the time of the primary eclipse, $T_0$, and the radius of the secondary, $R_2$. As free fitting parameters, we used the size of the line formation region of the respective line and the argument of periastron, as this can have an impact on the phasing of the timing of the eclipse. For the argument of periastron, we employed as a prior the posterior distribution obtained from the RV fit with the PHOEBE code (see Sect.~\ref{sec:PHOEBE}). For the likelihood function, we again used a least squares method.

    Since the \ion{N}{V} and \ion{C}{IV} resonance lines exhibit the same terminal wind velocity and the TVS spectra shown in Fig.~\ref{fig:TVS_NV} demonstrate comparable line formation profiles, we concluded that they are formed in the same region of the wind (see Sect.~\ref{sec:wind_variability_sub}). Therefore, we applied this method to estimate the size of their line formation region, using the information on their different spectral shapes as obtained from the HST spectra.

    In Fig.~\ref{fig:CIV_LC}, the phased projected distance between the primary's \ion{N}{V} and \ion{C}{IV} resonance line formation region and the secondary for our best-fitting model is displayed. One can see that, from our best-fitting model and its uncertainties, the projected distance does not show too much variation, with most of the variation originating from the uncertainty in the argument of the periastron. However, the instance when the secondary moves behind the primary's wind shows stronger variations and helps to limit the extent of the primary's wind. Following our best model, the measurable size of the resonance line formation regions in the primary's stellar wind is $R_\mathrm{1,\,wind}=316.3^{+12.4}_{-26.6}\,\rsun$ and the argument of periastron is $\omega=96.2^{+0.8}_{-1.5}\,^\circ$. The resulting posterior distributions are shown in Figs.~\ref{fig:CIVposterior}. The argument of the periastron is by a bit more than $1^\circ$ smaller than the one derived from the orbital analysis. We tested if the change, $\omega$, has a noticeable effect on the RV or LC fit, but could not see a significant change in the fit's quality.

    \section{Discussion}

    \subsection{Effects of the secondary's radiation field on the primary's stellar wind}
    \label{sec:discuss_variability}
    \begin{figure}
        \centering
        \begin{tikzpicture}[scale=1]
            \node[anchor=south] at (0, -1) {\includegraphics[trim= 1.05cm 2.2cm 1.7cm 2.8cm ,clip,width=\linewidth]{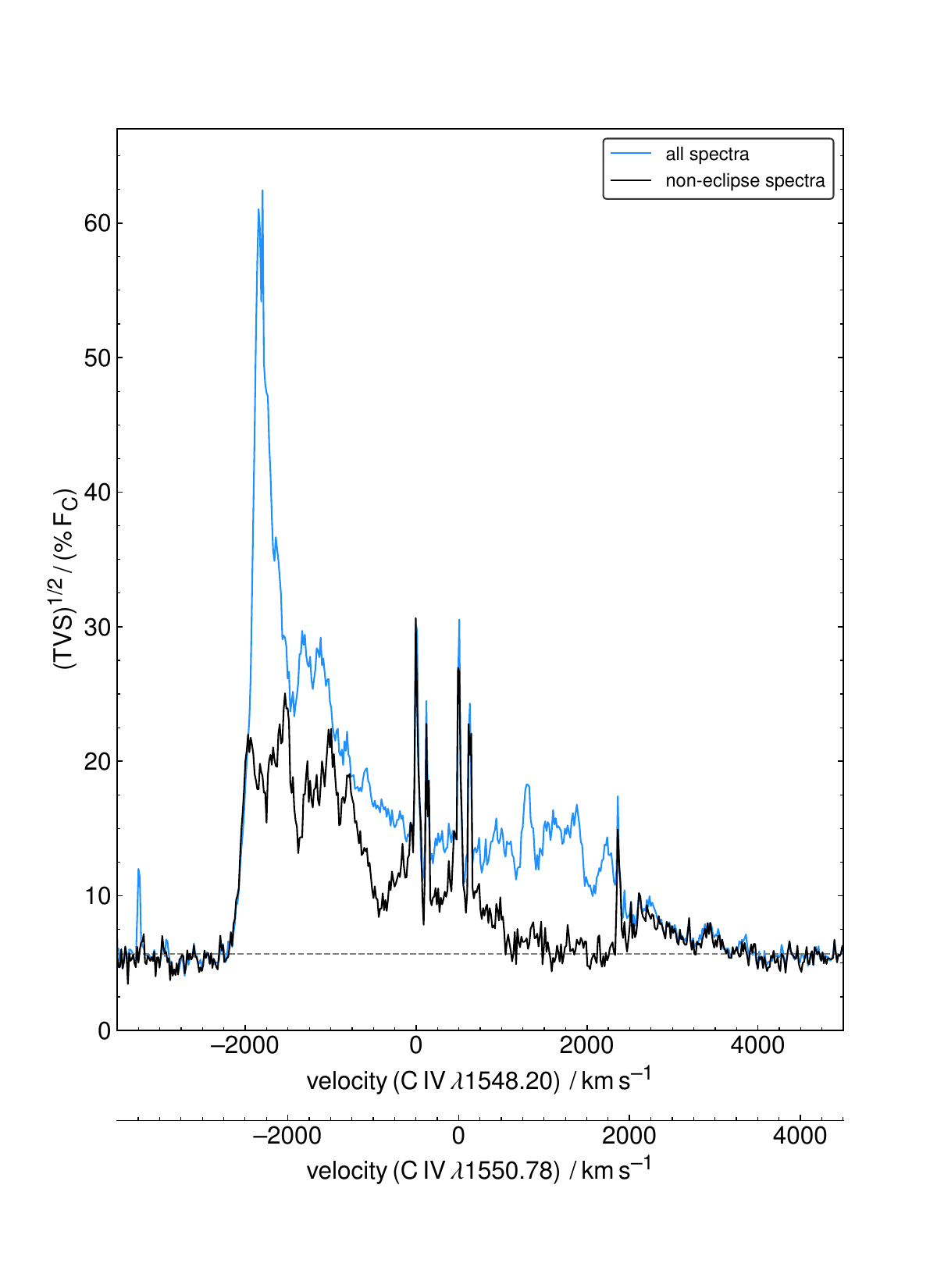}};
            \node at (0.5, 8.15-2.9) [draw, fill=white, minimum size=-0.2cm,white]{\fontsize{10}{16}\color{red}\selectfont change in emission};
            \node at (0.5, 7.7-2.9) [draw, fill=white, minimum size=-0.2cm,white]{\fontsize{10}{16}\color{red}\selectfont during wind eclipse};
            \draw[-latex,red,thick] (0.5,7.3-2.9) -- (0.5,6.3-2.9);
            \draw[red,very thick] (-0.2,5.8-2.9) -- (1.5,5.8-2.9);
            \draw[red,very thick] (-0.2,5.8-2.9) -- (-1.5,4.7);
            \node at (-0.5, 11.6-2.9) {\fontsize{10}{16}\color{blue}\selectfont variation in $\varv_\infty$,};
            \node at (-0.5, 11.15-2.9) {\fontsize{10}{16}\color{blue}\selectfont illumination};
            \node at (-0.5, 10.7-2.9) {\fontsize{10}{16}\color{blue}\selectfont by companion?};
            \draw[-latex,blue,thick] (-0.75,10.25-2.9) -- (-1.5,9.25-2.9);
            \node at (3.2, 6.75-2.6) [draw, fill=white, minimum size=-0.2cm,white]{\fontsize{10}{16}\color{blue}\selectfont additional emission,};
            \node at (3.2, 6.3-2.6) [draw, fill=white, minimum size=-0.2cm,white]{\fontsize{10}{16}\color{blue}\selectfont illumination};
            \node at (3.2, 5.85-2.6) {\fontsize{10}{16}\color{blue}\selectfont by companion?};
            \draw[-latex,blue,thick] (2.8,5.6-2.7) -- (2.3,5.2-2.7);
        \end{tikzpicture}    
        \caption{TVS in velocity space with respect to the \ion{C}{IV} $\lambda1548.20$ ($\lambda1550.70$, lower ticks) line. The TVS are presented in the full rest frame of the primary corrected for the systemic velocity. The TVS shown  in blue was calculated using all available spectra and the one in black using only the spectra where the secondary is not eclipsed by the wind (i.e. black spectra in Fig.~\ref{fig:PCygni}). The variance in the black TVS quickly drops off for redward velocities, while in the blueward \ion{C}{IV} absorption trough one can see some variance across the spectra, typical for a O-type star. In comparison, the blue TVS shows much stronger variance. The major contributor to the strong variance is the change in the \ion{C}{IV} when the primary's wind is eclipsing the secondary star. However, one can also see that there is enhanced variance at the edge of the \ion{C}{IV} absorption trough around $\approx\SI{-2000}{km\,s^{-1}}$ with respect to the \ion{C}{IV} $\lambda1548.20$ line, as well as some additional variance in the emission feature out to $\approx\SI{3000}{km\,s^{-1}}$ (equating therefore to $\approx\SI{2500}{km\,s^{-1}}$ in the rest frame of the \ion{C}{IV} $\lambda1550.70$ line).
        }
        \label{fig:TVS_full}
    \end{figure}
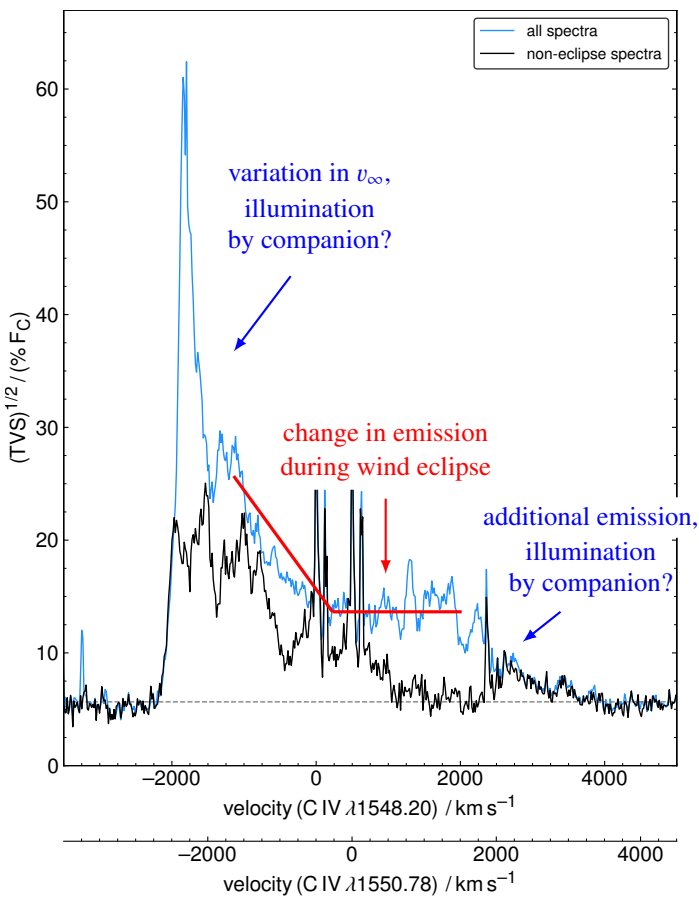

    The results presented above focus primarily on the dominant sudden drop in intensity in the observed stellar-wind lines of the early-type primary star. However, because the system contains two massive UV-bright stars, their respective winds and radiation fields may interact. Given that the wind of the secondary appears to be relatively weak, wind collisions are not expected. However a mutual irradiation could lead to a local heating of the stellar winds, and thus a change in the line shape.

    To search for additional variability in the primary’s wind induced by the secondary, first all UV spectra containing the \ion{C}{IV} resonance line were shifted to match the \ion{Si}{II} $\lambda\lambda 1526.72, 1533.45$ ISM lines. This should correct for potential wavelength-calibration offsets. Then each spectrum was shifted by the primary’s RV at the orbital phase of that observation to bring all data into the same rest frame (i.e. that of the primary star). The corresponding list of applied RV shifts for each spectrum is provided in Table~\ref{tab:CIV_RVs}. Lastly, these spectra were used to compute TVS of the \ion{C}{IV} line in order to search for variability caused by the secondary.

    Figure~\ref{fig:TVS_full} displays the calculated TVS in the rest frame of the \ion{C}{IV} $\lambda1548.20$ line for two cases: i) all available UV spectra, and ii) only the spectra in which the secondary is not eclipsed by the primary’s wind (i.e. the black spectra in Fig.~\ref{fig:PCygni}). When only the non-eclipse spectra are considered, the TVS shows the expected level of variability in the \ion{C}{IV} absorption trough, while the emission component remains largely stable, as is typical for O-type stars. However, when including all spectra, this enhances the variance across the profile noticeably. This is largely due to the sudden drop in \ion{C}{IV} line intensity discussed in Sect.~\ref{sec:wind_variability_sub}.

    In addition to this expected behaviour, further features emerge. A weak but noticeable variance appears in the emission component at redward velocities of approximately $\SIrange{2000}{2500}{km\,s^{-1}}$. This excess may originate from local wind heating caused by the secondary when it is at quadrature or partially eclipsed by the primary's wind, affecting the outer regions where the \ion{C}{IV} emission forms. Furthermore, a pronounced variance arises at the blue edge of the \ion{C}{IV} absorption trough, but only when considering the spectra in which the secondary becomes eclipsed by the primary's wind. While time series of wind-variable O-type stars commonly display variability at the blue edge of the \ion{C}{IV} absorption trough (often attributed to line deshadowing instabilities, e.g. \citealt{Owocki1984,Driessen2022b}), such strong and phase-dependent changes are atypical.

    \begin{figure}
    \centering
    \includegraphics[trim= 1.1cm 0.8cm 2.7cm 1.9cm ,clip,width=\linewidth]{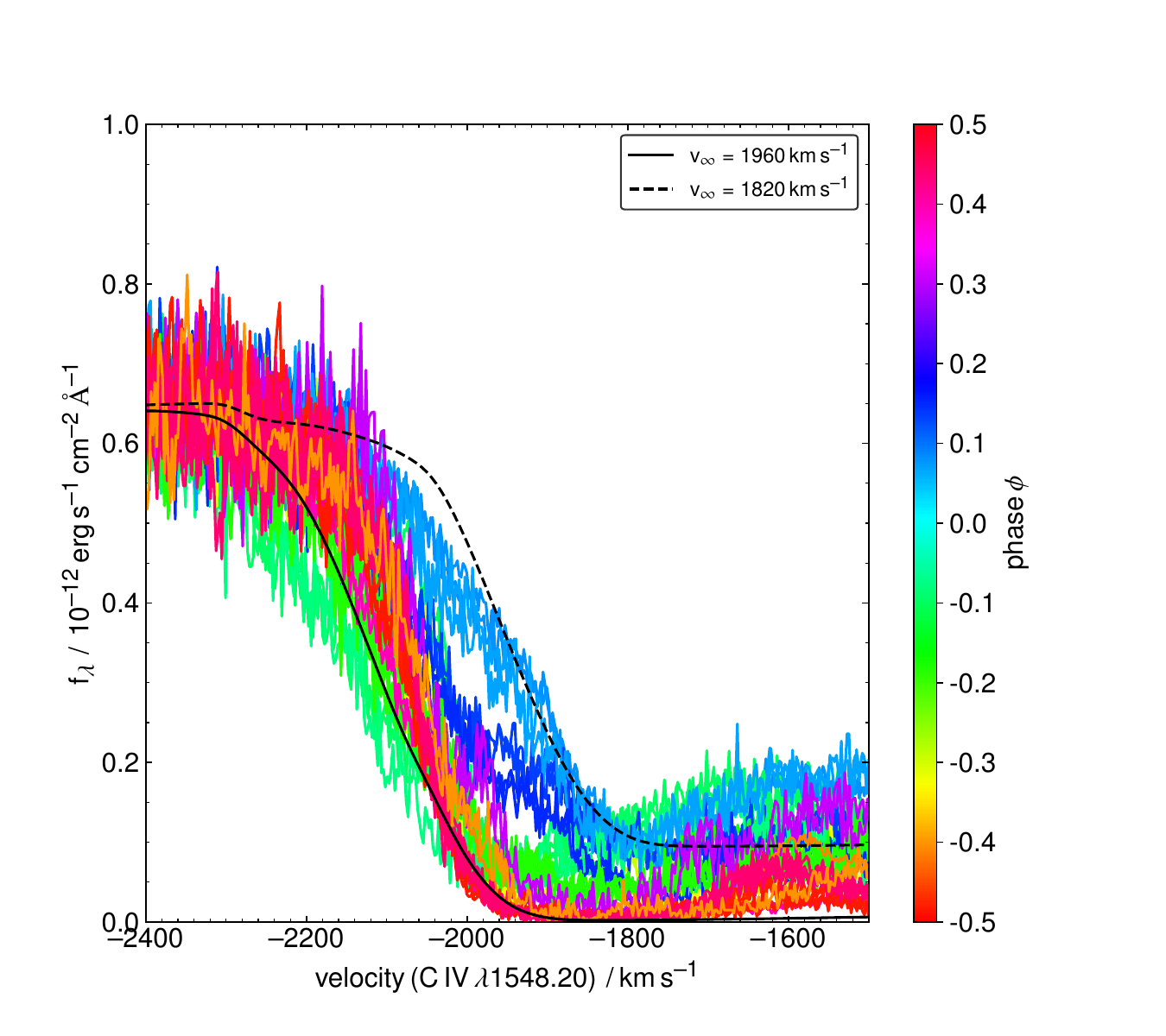}
        \caption{Zoom-in on the blue edge of the \ion{C}{IV} absorption trough in velocity space with respect to the \ion{C}{IV} $\lambda1548.20$ line. All spectra are shown in the frame of rest of the primary star and are corrected for the systemic velocity. For clarity only high-resolution spectra are shown. The spectra are colour‑coded by orbital phase. The black lines show synthetic \ion{C}{IV} resonance line profiles using different stellar wind velocities from the primary and considering no (solid) or some contribution of the secondary (dashed). 
        }
        \label{fig:vinf_edge}
    \end{figure}

    Figure~\ref{fig:vinf_edge} presents a zoom‑in on the blue edge of the \ion{C}{IV} absorption trough for all available spectra. The blue edge remains remarkably constant in the majority of the spectra, namely in the spectra taken at phases at which the secondary is being eclipsed by the primary's wind, as well as phases at which the secondary and primary are physically distant (i.e. $\phi\approx\SIrange{-0.5}{-0.15}{}$ and $\phi\approx\SIrange{0.15}{0.5}{}$). In contrast, spectra taken close to the primary eclipse (i.e. when the stars are physically closest; $\phi\approx\SIrange{-0.1}{0.1}{}$) display clear deviations: the blue edge shifts by several hundred kilometres per second and its slope changes significantly. Interestingly, the blue edge of the \ion{C}{IV} absorption trough is displaced bluewards before the eclipse and redwards afterwards, whereas the change in slope appears to be similar in both cases. 
    
    Figure~\ref{fig:orbit} illustrates the orbital configuration at different orbital phases. At $\phi = -0.1$, the secondary is still located outside the primary’s \ion{C}{IV} line‑forming region. Nonetheless, at this phase the emergent flux from the secondary starts to exceed the flux from the primary in the outer parts of the \ion{C}{IV} resonance line‑formation zone. The additional incoming radiation from the secondary likely heats these outer layers, increasing macro-turbulent motions and thereby altering the slope of the blue edge such as seen by the behaviour in the green spectra in Fig.~\ref{fig:vinf_edge}.

    As the system approaches periastron (given the geometry of the system, this is very shortly after the primary eclipse), the secondary moves inside the \ion{C}{IV} line‑formation region of the primary's wind along the observer’s line of sight. Although the secondary is much smaller in physical size, preventing the secondary itself from significantly blocking the primary’s flux, its ionizing radiation can still affect a substantial fraction of the surrounding wind. Because this irradiation impacts the outer regions where the \ion{C}{IV} absorption is formed, it should produce an apparent shift in the blue edge of the trough.

    Unfortunately, no HST spectra have yet been obtained exactly during this critical phase. The wind relaxation timescale is about a few days, meaning the resulting perturbations can only be observed in spectra taken while the secondary is within the primary's wind or shortly afterwards. The closest post‑periastron spectrum that is available occurs at $\phi = 0.08$, approximately $\SI{12}{d}$ after periastron and about $\SI{5}{d}$ after the secondary has left the \ion{C}{IV} line‑formation region. This spectrum is shown in light blue in Fig.~\ref{fig:vinf_edge} and indeed exhibits a blue edge corresponding to an apparently reduced terminal wind velocity and a shallower slope. This is both consistent with the effect of the secondary’s irradiation onto the primary's stellar wind.

    The next available HST observation, taken at $\phi = 0.15$ (dark blue in Fig.~\ref{fig:vinf_edge}), reflects a configuration where the stars are again well separated, such that the secondary no longer influences the primary’s wind. Correspondingly, the blue edge has nearly returned to its unperturbed shape, matching the spectra obtained at other unaffected phases.

    Despite the strong indications that the secondary does influence the primary’s wind structure close to the periastron passage, these effects appear to be of lesser importance compared to the dramatic drop in the \ion{C}{IV} P\,Cygni profile occurring when the primary wind fully eclipses the secondary star. Given that we do not rely on the change in the blue edge of the \ion{C}{IV} absorption trough or the extension in the emission feature, we doubt that these effects will have a significant impact on the measurement of the line formation region. 
       
    \subsection{Searching for imprints of the secondary’s wind in the \ion{C}{IV} resonance line}
    \label{sec:secondary_wind}

    Although the secondary appears to have a weak wind, it may still contribute subtle wind signatures embedded in the UV resonance lines that could provide constraints on its properties. To search for such contributions, we first assumed that the secondary possesses a faster wind than the primary ($\varv_{\infty,2} > \varv_{\infty,1}$). In the case of a weak wind, this would result in an additional \ion{C}{IV} absorption component at bluer wavelengths, producing a shallow, step-like decrease in the continuum at the blue edge of the \ion{C}{IV} resonance-line absorption trough. However, as shown in Fig.~\ref{fig:PCygni}, no such change in intensity is observed. Furthermore, given a peak-to-peak radial-velocity amplitude of approximately $\SI{200}{km\,s^{-1}}$ for the secondary, this feature should be readily detectable as a clear shift when viewed in the reference frame of the primary. However, as demonstrated in Fig.~\ref{fig:vinf_edge}, no variability is detected at the blue edge, thereby ruling out the presence of a detectable fast wind from the secondary.

\begin{figure}
    \centering
    \includegraphics[trim= 0.2cm 0.8cm 1.2cm 1.9cm ,clip,width=\linewidth]{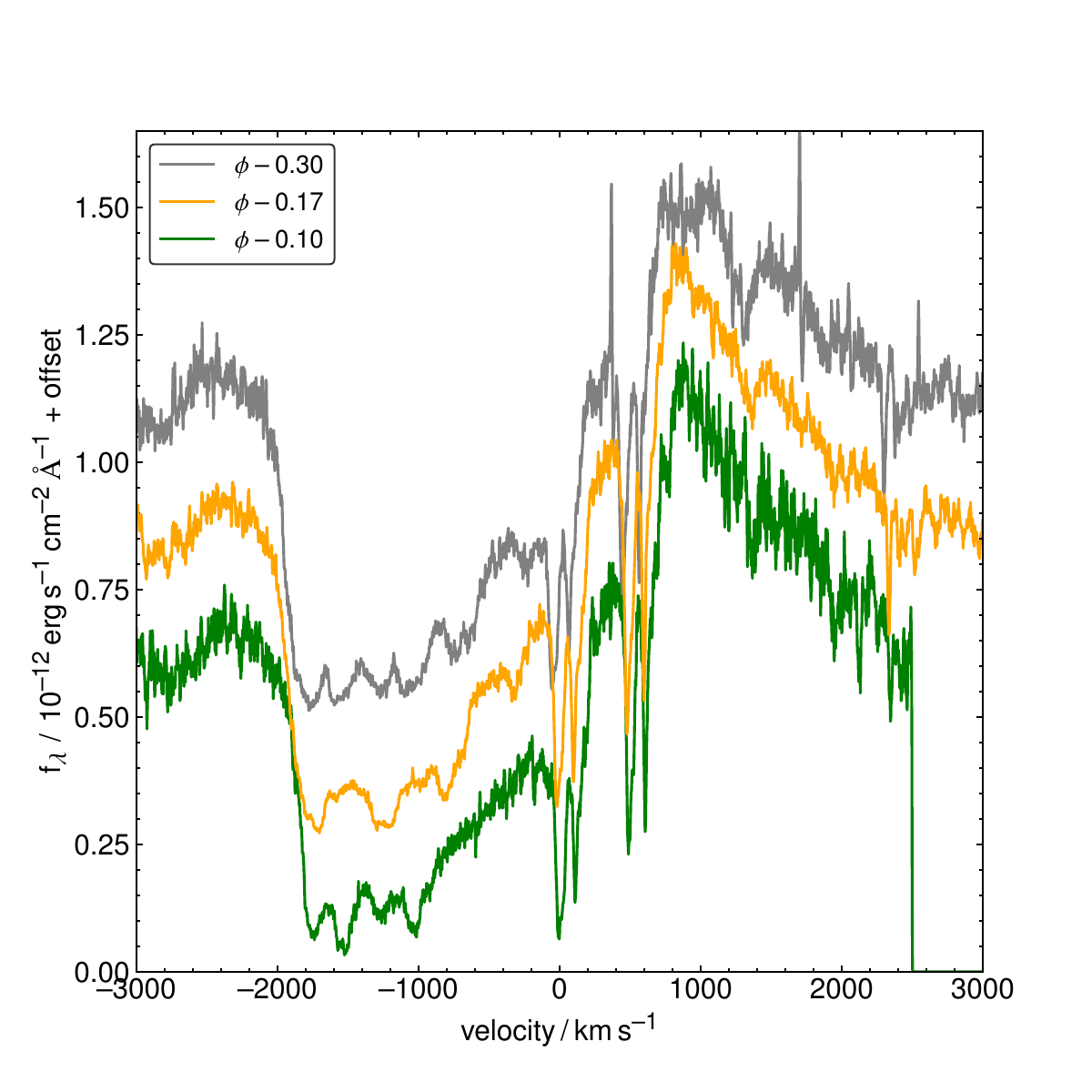}
        \caption{Detail of the \ion{C}{IV} resonance line profiles of selected spectra between phases $\phi = -0.3\text{\,--\,}-\!0.1$ shifted into the rest frame of the secondary star and corrected for the systemic velocity. For clarity, the spectra at phases $\phi=-0.3$ (grey) and $\phi=-0.17$ (orange) are shifted by $\SI{0.5e-12}{erg\,s^{-1}\,cm^{-2}\,\AA^{-1}}$ and $\SI{0.25e-12}{erg\,s^{-1}\,cm^{-2}\,\AA^{-1}}$, respectively. One can see narrow absorption features at velocities of around $\SI{-150}{km\,s^{-1}}$, $\SI{-850}{km\,s^{-1}}$, $\SI{-1000}{km\,s^{-1}}$, and $\SI{-1500}{km\,s^{-1}}$ emerging and disappearing. Note that the NACs around $\SI{-1300}{km\,s^{-1}}$ and $\SI{-1800}{km\,s^{-1}}$ are attributable to the primary star's wind, and therefore, as expected, move when plotted in the rest frame of the secondary.
        }
        \label{fig:nac_features}
    \end{figure}

    As a second possibility, we assumed that the secondary has a slower wind than the primary ($\varv_{\infty,2} < \varv_{\infty,1}$). Although weak profile variations in a P\,Cygni profile are difficult to detect, signatures of narrow absorption components (NACs, see \citealt{Prinja1986,Howarth1989}) forming in the outermost regions of the secondary’s wind might still be observable. We therefore inspected all available UV spectra and found that the observations obtained between orbital phases $\phi = -\,0.3\text{\,--\,}-0.1$ exhibit peculiar, narrow low-velocity absorption features (see Fig.~\ref{fig:nac_features}). To test whether these features are associated with the secondary, we shifted all spectra containing these features into the reference frame of the secondary star. If the features were intrinsically linked to the secondary, they should align in velocity space. As shown in Fig.~\ref{fig:nac_features}, this alignment is not observed, ruling out an origin as NACs in the secondary’s wind. We conclude that these low-velocity absorption features must instead be associated with additional, local optically thick structures in the primary’s wind. It remains unclear whether these features are indirectly linked to the presence of the secondary interacting with the primary's wind or represent time-dependent phenomena, such as localized bright spots on the primary that generate additional co-rotating interaction regions \citep{Owocki1995,Cranmer1996,Prinja2002,David2017}.

   As a third option, we investigated the possibility that the secondary star also has a stellar wind and that wind directly produces the variations in the slope of the blue C IV edge seen in Fig.~\ref{fig:vinf_edge}, rather than these being due to additional heating of the secondary star's radiation field as discussed above. If the change in slope of the blue edge observed close to the periastron passage ($\phi=0$) were due to the wind of the secondary star, this would require the wind of the secondary star to coincidentally have the same terminal velocity as the primary ($\varv_{\infty,1} \approx \varv_{\infty,2}$).
    
    This scenario can be tested by considering the blue edge slopes in more detail in the spectra where the two stars are approaching each other. As these spectra are hard to distinguish in Fig.~\ref{fig:vinf_edge}, we show in Fig.~\ref{fig:vinf_edge_selected} the blue edge shapes at $\phi = -0.18$ and $\phi = -0.1$, each compared to a reference spectrum where the stars are physically well separated ($\phi = -0.48$). Given that the RV separation of the secondary to the primary is very similar at both of these phases, a similar change in the slope of the blue edge resulting from a putative wind of the secondary star should be seen. As Fig.~\ref{fig:vinf_edge_selected} demonstrates, this effect is not observed. In addition, if the changes to the blue absorption edge were considered to be a signature of the secondary star's wind alone, that would not explain the simultaneous changes in the emission profiles (see Fig.~\ref{fig:PCygni}).

    While not entirely discounting the possibility that the secondary star's wind signature may make a contribution to the changes in the observed absorption blue edge, we repeated the calculations for the size of the primary's line formation region described in Section 3.2 assuming a strong, extended secondary wind. Even in the extreme case of a line-forming region extending out to 10 times the stellar radius (i.e. $10R_2\approx 100\,\rsun$), the extent of the primary star's wind line-forming region would increase only from $316\,\rsun$ to approximately $400\,\rsun$. Note that for more extended winds one would expect that these winds collide and produce X-rays.
    
    \subsection{Comparison to stellar atmosphere models}
    
    \begin{figure}
        \centering
        \begin{tikzpicture}
            \node[anchor=south] at (0, 6) {\includegraphics[trim= 1.cm 0.2cm 2cm 1.3cm ,clip,width=\linewidth]{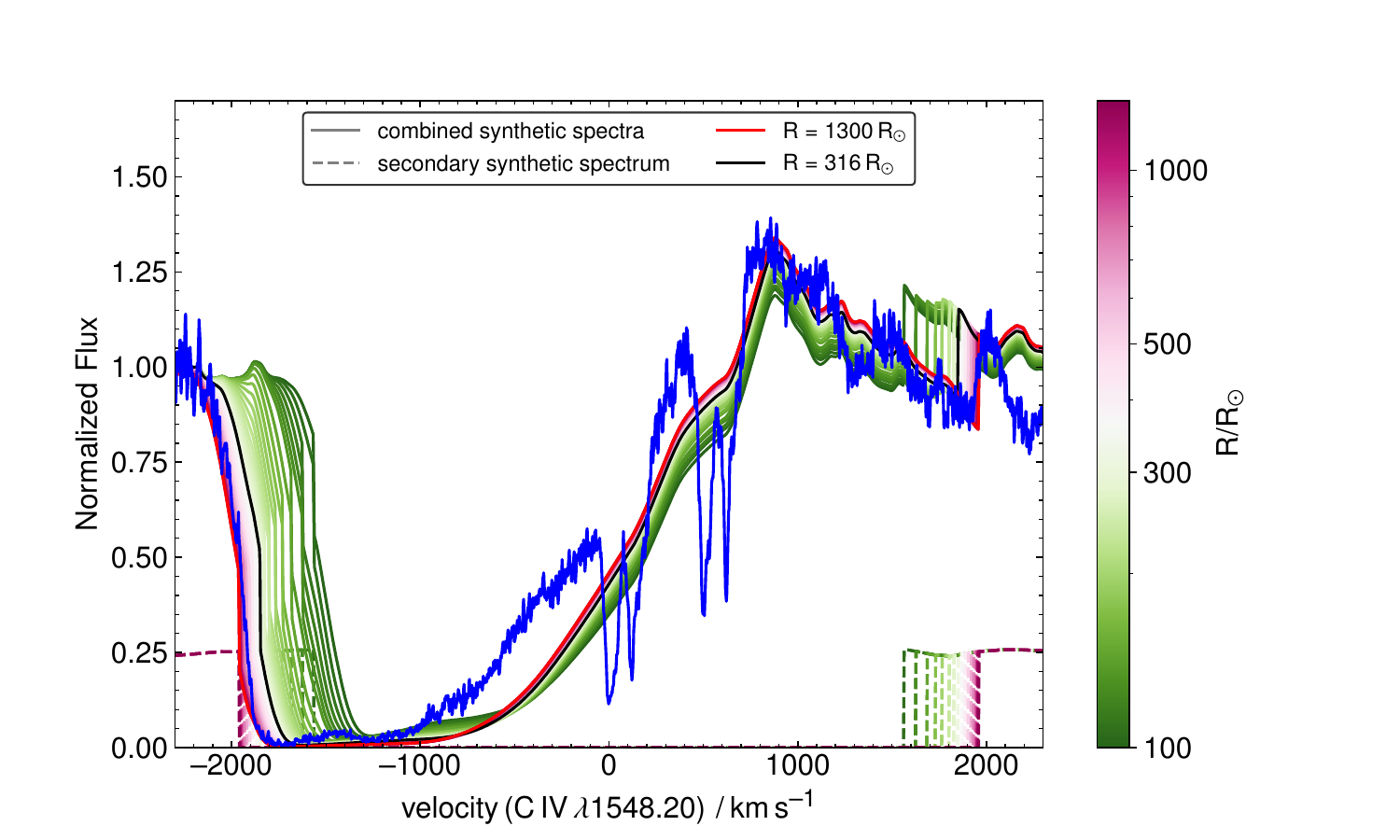}};
            \node[fill=white, inner sep=2pt] at (-0.3, 11.7) {$\beta=0.8$}; 
            \node[anchor=south] at (0, 0) {\includegraphics[trim= 1.cm 0.2cm 2cm 1.3cm ,clip,width=\linewidth]{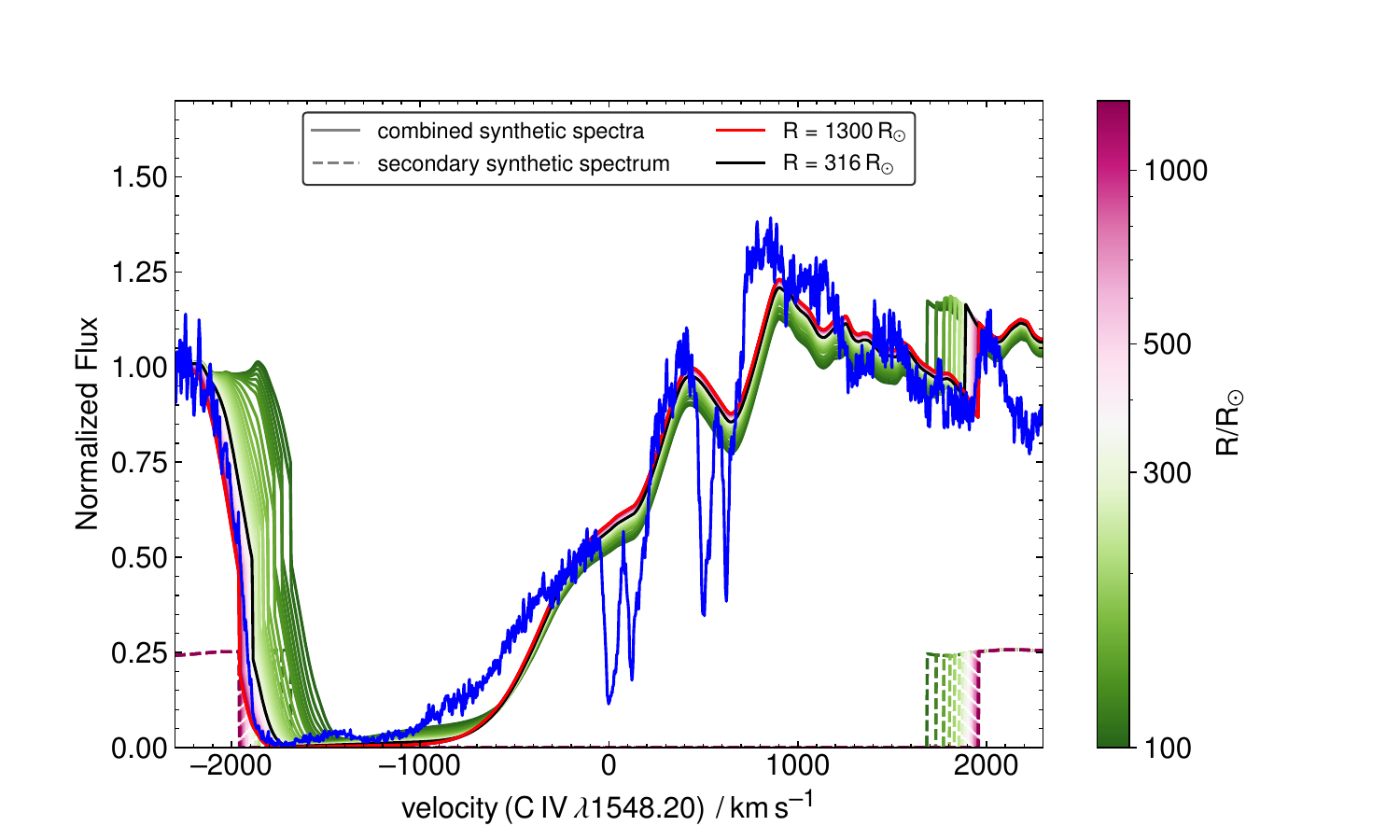}};
            \node[fill=white, inner sep=2pt] at (-0.3, 5.7) {$\beta=0.5$}; 
        \end{tikzpicture}
        \caption{Comparison between the observed \ion{C}{IV} P\,Cygni profile obtained near phase $\phi\approx0.5$ (blue) and combined synthetic spectra of the primary and secondary for different assumed radial extents of the primary’s stellar wind (colour‑coded lines). At this orbital phase, the \ion{C}{IV} resonance line contains only flux originating from the primary’s wind, whereas the observed continuum is the sum of the flux contributions from both stars. To account for this effect in a simple way, the secondary’s continuum flux is added to the primary spectrum once the terminal wind velocity is reached. For clarity, only the secondary’s contribution to the combined spectrum is shown as dashed colour‑coded lines. The synthetic profile corresponding to the largest wind extent ($\sim1300\,R_\odot$) is shown as a solid red line, while the profile computed up to the measured radius of $316\,R_\odot$ is shown as a solid black line. The wind velocity field of the primary in the two panels is described by a $\beta$ law with exponent $\beta = 0.8$ (upper panel) and $\beta = 0.5$ (lower panel).
              }
        \label{fig:profiles}
   \end{figure}

    In principle, each spectrum received by the observer should be computed by fully accounting for the orbital configuration and the mutual interaction between the stellar atmospheres and radiation fields. However, this is only possible when using multi‑dimensional simulations, which are to date computationally too expensive even for a single star's full atmosphere. Even in lower dimensions, no such framework currently exists. Fortunately, the observational evidence indicates that the two stars only affect each other’s atmospheres during a limited fraction of the orbit (see Sect.~\ref{sec:discuss_variability}). For most orbital phases, 1D stellar atmosphere models therefore provide a reasonable approximation of each star's stellar atmosphere.

    The 1D stellar atmosphere grid models used in this work to roughly estimate the stellar parameters of the primary also successfully reproduce the overall shape and strength of the \ion{C}{IV} resonance line. To test whether they can also explain the observed spatial extent of the line‑formation region, we adjusted the terminal wind velocity of the primary's model to $\varv_{\infty,1} = \SI{1960}{km\,s^{-1}}$ while using a classical $\beta$ law with exponent $\beta=0.8$ typical for O-stars \citep{Pauldrach1986,Puls1996}. For a comparison of the model to the observed spectra, those spectra taken around phase $\phi\approx0.5$ are particularly well suited, since at this phase the \ion{C}{IV} resonance line originates exclusively in the primary’s wind, while the continuum flux still contains contributions from both stars.

    To account for the contribution of the secondary to the combined spectrum at this configuration, the secondary’s continuum flux was added to the primary spectrum only outside the velocity range in which the primary’s \ion{C}{IV} absorption was formed. In the stellar atmosphere models, the wind structure was computed out to $\approx1300\,R_\odot$ (i.e. $50\,R_\star$). To determine the outer limit of the wind region in which the P\,Cygni profile is formed, we computed synthetic spectra of the primary by truncating the wind at selected radii. 
    
    The resulting profiles are depicted in the upper panel of Fig.~\ref{fig:profiles}.  In the model, \ion{C}{IV} remains the dominant ionization stage throughout the wind, causing the absorption to continue building up until the outer boundary is reached. This can be seen by the \ion{C}{IV} resonance line building up until the terminal velocity at radii of order $\sim1300\,R_\odot$ is reached. When comparing the observed spectrum where the secondary is eclipsed by the primary's wind with the combined synthetic \ion{C}{IV} profile produced using a wind truncated at $316\,R_\odot$, it becomes evident that the model fails to reproduce the observed line shape as it is formed only at much lower wind velocities. This brings the adopted velocity stratification into question. 

    Another issue with this model is that the shape of the P\,Cygni profile is not well matched and that the absorption trough in the model spectrum is too deep at low velocities. Although the irradiation of the secondary can in principle have an effect on the line shape, it should not be sufficient to lift this degeneracy.

    With a changed velocity stratification, this issue might be resolvable. From line‑driven wind theory (known as CAK theory; \citealt{Castor1975}), a lower exponent for the wind velocity law of $\beta = 0.5$ is predicted. With this exponent, the terminal wind velocity will be asymptotically approached closer to the photosphere. This should bring the model spectrum closer to the observationally inferred line formation radius.

    The lower panel of Fig.~\ref{fig:profiles} shows the resulting \ion{C}{IV} line profiles for the updated primary's stellar atmosphere model with $\beta = 0.5$. As expected, the wind accelerates more rapidly and the terminal velocity is approached much deeper in the wind, with velocity flattening in the outer regions. The main formation region of the \ion{C}{IV} resonance line is now confined to the inner wind, within the first few hundred solar radii, in much better agreement with the observations. Moreover, the width the absorption trough is better reproduced, further supporting a steeper velocity law.

    Nevertheless, these results should be interpreted with caution. The strength and shape of the \ion{C}{IV} resonance line depend not only on the wind velocity field, but also on the assumed mass‑loss rate, temperature structure, and chemical abundances (most notably the surface carbon abundance). To robustly benchmark stellar atmosphere models for this system, multi‑epoch optical spectroscopy is required to reliably constrain temperatures and abundances of the two stars. Combined with the UV data, this would enable a reliable determination of the primary’s wind velocity structure to a so-far unreached accuracy. 

\section{Conclusions}

    In this work, we present the first measurement of the extent of UV resonance line formation regions in the stellar wind of a O-type star, using the primary of the SMC eclipsing binary system AzV~75. This is achieved by combining the high-quality photometric observations in the g band from the ASAS-SN and the TESS survey with RV measurements of both binary companions from HST UV spectra. 

    The orbital analysis revealed that the binary has a long period of $P=\SI{165.925}{d}$ and that the binary components are on quite an eccentric orbit, with $e=0.48$. The two stars have masses of ${M_1=54\,\msun}$ and $M_2=41.5\,\msun$, making them among the most massive stars in the SMC. From the light curve modelling, we extracted a radius of $R_1=25.5\,\rsun$ for the primary, consistent with its classification as a giant, a radius of $R_2=9.65\,\rsun$ for the secondary, consistent with a main-sequence companion.

    The UV resonance lines in the HST spectra show strong variations in intensity with three different states: i) maximum emission and in-filled absorption trough, ii) a $\approx20\%$ drop in flux in the emission feature and black absorption trough, and iii) an intermediate stage. All these different stages are strongly correlated with the orbital phase. We illustrate that this behaviour can be explained by the primary's stellar wind eclipsing the secondary. Such wind eclipses only block the light at optically thick wavelengths (i.e. the resonance lines), with the continuum remaining unaffected. Using this effect in combination with multi-epoch HST data, we were able to derive the extent of the line formation regions of the primary star.

    Both the \ion{N}{V} and \ion{C}{IV} resonance lines reach a terminal wind velocity of $\varv_\infty=\SI{1960}{km\,s^{-1}}$. Therefore, we assume that these lines are formed over the full extent of the wind, and that we can measure the actual size of their line formation regions, and thus the size of the primary's stellar wind. Using an analytic model, we were able to derive that the line formation region of the \ion{N}{V} and \ion{C}{IV} resonance lines in the primary's wind extend to $R_\mathrm{1,\,wind}=316.3\,\rsun$.

    From an inspection of the blue edge of the \ion{C}{IV} absorption trough of all spectra across the different orbital phases covered by the HST spectra, we could detect that the secondary star interacts with the primary's stellar wind at phases around $\phi\approx\SIrange{-0.1}{0.1}{}$. At first, the additional UV radiation of the secondary leads to a heating of the stellar wind and introduces large-scale turbulence. Then, at the periastron passage, the secondary passes through the primary's wind, causing additional heating and turbulence resulting in an apparent reduction in the terminal wind velocity. Given the short dynamical timescale of the primary's stellar wind, it recovers quickly from this perturbation and remains at its full extent for the rest of the orbital phases. 
    
    Using the information on the stellar masses, radii, and the total luminosity of the system, we were able to find approximate stellar atmosphere models using the publicly available PoWR OB-Vd3 grids. We updated the terminal wind velocity to match the observations and find that the measured size of the \ion{C}{IV} resonance line formation region can only be reproduced when assuming a $\beta$ law with exponent $\beta=0.5$, thereby providing new, independently derived constraints on the velocity structure of massive star winds.

\begin{acknowledgements}
      DP acknowledges financial support from the FWO in the form of a junior postdoctoral fellowship No. 1256225N. This research is based on observations made with the NASA/ESA Hubble Space Telescope obtained from the Space Telescope Science Institute, which is operated by the Association of Universities for Research in Astronomy, Inc., under NASA contract NAS 5–26555. This paper includes data collected by the TESS mission, which are publicly available from the Mikulski Archive for Space Telescopes (MAST). Funding for the TESS mission is provided by the NASA's Science Mission Directorate. The authors acknowledge the data provided by the ASAS-SN project, which is funded by the Gordon and Betty Moore Foundation and supported by the Las Cumbres Observatory.
\end{acknowledgements}

\bibliographystyle{aa}
\bibliography{aa60585-26}

\begin{appendix} %

\onecolumn
\section{Additional tables}

   \begin{table}[h]
      \caption[]{List of spectra from HST UV observations of AzV 75 used for the RV measurements.}
         \label{tab:RVs}
        \centering
        \begin{tabular}{lcclccccc}
        \hline
        \hline
        Proposal    & Spec. ID  & Instrument & Grating           & \multicolumn{1}{c}{Start Time}   & Phase             & $\varv_\mathrm{corr}^\dagger$        & RV$_1$                              & RV$_2$                               \\
                    &           &            &                   & \multicolumn{1}{c}{[JD-2400000.5]}&                  & \multicolumn{1}{c}{[km\,s$^{-1}$]}  & [km\,s$^{-1}$]                      & [km\,s$^{-1}$]                       \\
        \hline\rule{0cm}{2.2ex}%

        7437        & o4wr11020 & STIS       & E140M             & 51322.323                        & \phantom{0}0.307  & \phantom{0}$0.0\pm0.3$               & \phantom{0}$84.8\pm17.5$            & $168.6\pm12.6$                       \\
        12805       & lbxk01010 & COS        & G130M             & 56134.870                        & \phantom{0}0.304  & \phantom{0}$1.5\pm0.6$               & \phantom{0}$77.5\pm\phantom{0}4.9$  & $163.7\pm13.3$                       \\
        12796       & lbxk51010 & COS        & G130M             & 56181.870                        & -0.413            & \phantom{0}$0.3\pm0.6$               & $140.9\pm\phantom{0}5.2$            & \phantom{0}$83.0\pm\phantom{0}9.8$   \\
        13122       & lc6a01010 & COS        & G130M             & 56366.959                        & -0.297            & $-0.1\pm0.7$                        & $162.8\pm10.0$                      & \phantom{0}$64.5\pm21.1$             \\
        13122       & lc6a01020 & COS        & G130M             & 56367.014                        & -0.297            & \phantom{0}$0.3\pm0.7$               & $167.4\pm\phantom{0}9.5$            & \phantom{0}$57.9\pm19.9$             \\
        13522       & lcgh01030 & COS        & G130M             & 56731.595                        & -0.100            & $-0.1\pm0.6$                         & $200.7\pm\phantom{0}6.6$            & \phantom{0}$17.7\pm22.4$             \\
        13522       & lcgh01040 & COS        & G130M             & 56731.606                        & -0.100            & $-0.7\pm0.6$                         & $207.5\pm\phantom{0}9.6$            & \phantom{0}$13.6\pm26.9$             \\
        13931       & lcrs01010 & COS        & G130M             & 57093.129                        & \phantom{0}0.078  & $23.6\pm0.6$                         & \phantom{0}$32.1\pm\phantom{0}7.2$  & $235.2\pm27.6$                       \\
        13931       & lcrs01030 & COS        & G130M             & 57093.215                        & \phantom{0}0.079  & $28.3\pm0.6$                         & \phantom{0}$32.3\pm\phantom{0}4.1$  & $251.9\pm19.5$                       \\
        13931       & lcrs51010 & COS        & G130M             & 57216.434                        & -0.179            & \phantom{0}$1.7\pm0.6$               & $192.7\pm14.5$                      & \phantom{0}$26.3\pm29.3$             \\
        13931       & lcrs51030 & COS        & G130M             & 57216.505                        & -0.178            & \phantom{0}$5.7\pm0.6$               & $195.8\pm\phantom{0}8.6$            & \phantom{0}$21.6\pm21.4$             \\
        14437       & ld2g01030 & COS        & G130M             & 57464.946                        & \phantom{0}0.318  & $64.1\pm0.6$                         & \phantom{0}$80.5\pm\phantom{0}7.3$  & $171.8\pm16.9$                       \\
        14437       & ld2g01040 & COS        & G130M             & 57464.956                        & \phantom{0}0.318  & $66.6\pm0.6$                         & \phantom{0}$88.6\pm13.6$            & $185.6\pm20.3$                       \\
        14437       & ld2g51030 & COS        & G130M             & 57562.012                        & -0.097            & \phantom{0}$0.5\pm0.6$               & $207.0\pm11.6$                      & \phantom{0}$19.7\pm23.9$             \\
        14437       & ld2g51060 & COS        & G130M             & 57562.075                        & -0.097            & \phantom{0}$2.9\pm0.6$               & $210.1\pm10.9$                      & \phantom{0}$19.3\pm27.1$             \\
        14842       & ldb301060 & COS        & G130M             & 57629.894                        & \phantom{0}0.312  & $12.6\pm0.8$                         & \phantom{0}$82.3\pm\phantom{0}6.5$  & $165.9\pm18.1$                       \\
        14842       & ldb301060 & COS        & G130M             & 57629.955                        & \phantom{0}0.313  & \phantom{0}$9.2\pm0.8$               & \phantom{0}$82.4\pm\phantom{0}5.1$  & $179.3\pm18.2$                       \\
        14855       & ldc301030 & COS        & G130M             & 57830.489                        & -0.479            & \phantom{0}$3.0\pm0.6$               & $132.3\pm13.9$                      & $100.3\pm26.4$                       \\
        14855       & ldc301040 & COS        & G130M             & 57830.500                        & -0.479            & \phantom{0}$4.8\pm0.6$               & $131.7\pm\phantom{0}8.3$            & \phantom{0}$99.5\pm26.3$             \\
        14909       & ldej3a010 & COS        & G130M             & 57847.572                        & -0.376            & $-0.1\pm0.7$                         & $152.6\pm14.5$                      & \phantom{0}$73.9\pm17.9$             \\
        14909       & ldej3a020 & COS        & G130M             & 57847.607                        & -0.376            & $-2.0\pm0.7$                         & $157.2\pm\phantom{0}8.2$            & \phantom{0}$76.9\pm27.5$             \\
        14909       & ldej3a030 & COS        & G130M             & 57847.635                        & -0.376            & \phantom{0}$1.2\pm0.6$               & $151.3\pm\phantom{0}8.0$            & \phantom{0}$83.1\pm20.5$             \\
        14909       & ldej3a040 & COS        & G130M             & 57847.674                        & -0.376            & \phantom{0}$2.7\pm0.7$               & $153.5\pm11.8$                      & \phantom{0}$67.8\pm26.3$             \\
        15385       & ldq701050 & COS        & G130M             & 58189.212                        & -0.318            & \phantom{0}$2.7\pm0.7$               & $158.4\pm\phantom{0}8.6$            & \phantom{0}$52.1\pm21.2$             \\
        15385       & ldq701080 & COS        & G130M             & 58189.333                        & -0.317            & \phantom{0}$3.4\pm0.7$               & $159.0\pm10.2$                      & \phantom{0}$57.9\pm22.2$             \\
        15536       & lduh01050 & COS        & G130M             & 58561.313                        & -0.076            & $-2.9\pm0.7$                         & $201.2\pm\phantom{0}9.9$            & \phantom{0}$13.7\pm27.7$             \\
        15536       & lduh01080 & COS        & G130M             & 58561.340                        & -0.076            & \phantom{0}$0.5\pm0.8$               & $194.8\pm10.9$                      & \phantom{0}$17.6\pm20.7$             \\
        15536       & lduh51050 & COS        & G130M             & 58659.839                        & -0.482            & $-3.3\pm0.7$                         & $135.7\pm\phantom{0}8.5$            & \phantom{0}$87.7\pm22.0$             \\
        15536       & lduh51080 & COS        & G130M             & 58659.865                        & -0.482            & \phantom{0}$0.5\pm0.8$               & $125.8\pm14.0$                      & $110.4\pm29.1$                       \\
        15774       & le0r01050 & COS        & G130M             & 58928.472                        & \phantom{0}0.137  & $-1.8\pm0.7$                         & \phantom{0}$29.5\pm\phantom{0}8.7$  & $226.5\pm24.1$                       \\
        15774       & le0r01080 & COS        & G130M             & 58928.527                        & \phantom{0}0.137  & \phantom{0}$1.7\pm0.7$               & \phantom{0}$31.7\pm\phantom{0}4.8$  & $225.4\pm30.9$                       \\
        16467       & leij01010 & COS        & G130M             & 59288.359                        & \phantom{0}0.305  & $-0.3\pm0.7$                         & \phantom{0}$80.3\pm10.7$            & $171.8\pm18.0$                       \\
        16467       & leij01030 & COS        & G130M             & 59288.429                        & \phantom{0}0.305  & \phantom{0}$0.6\pm0.7$               & \phantom{0}$76.1\pm\phantom{0}6.7$  & $170.7\pm16.6$                       \\
        16467       & leij01040 & COS        & G130M             & 59288.464                        & \phantom{0}0.305  & \phantom{0}$4.1\pm0.6$               & \phantom{0}$76.3\pm\phantom{0}5.6$  & $161.1\pm16.9$                       \\
        16325       & lefc01050 & COS        & G130M             & 59414.851                        & \phantom{0}0.067  & $-2.0\pm0.7$                         & \phantom{0}$32.1\pm10.3$            & $238.4\pm26.4$                       \\
        16325       & lefc01080 & COS        & G130M             & 59414.878                        & \phantom{0}0.067  & \phantom{0}$2.3\pm0.8$               & \phantom{0}$22.9\pm12.6$            & $219.8\pm32.7$                       \\
        16534       & leko01050 & COS        & G130M             & 59760.728                        & \phantom{0}0.151  & $-2.5\pm0.7$                         & \phantom{0}$38.3\pm\phantom{0}4.8$  & $226.7\pm17.9$                       \\
        16534       & leko01080 & COS        & G130M             & 59760.754                        & \phantom{0}0.151  & \phantom{0}$1.1\pm0.7$               & \phantom{0}$54.7\pm25.4$            & $229.1\pm29.6$                       \\
        17250       & lf2301070 & COS        & G130M             & 60140.727                        & \phantom{0}0.441  & $-4.8\pm0.7$                         & $107.2\pm10.3$                      & $127.1\pm21.5$                       \\
        17250       & lf23010a0 & COS        & G130M             & 60140.779                        & \phantom{0}0.441  & $-4.5\pm0.8$                         & $109.2\pm\phantom{0}5.0$            & $144.6\pm28.6$                       \\
        17327       & lf4p01070 & COS        & G130M             & 60499.317                        & -0.399            & $56.4\pm2.3$                         & $149.6\pm15.2$                      & \phantom{0}$77.2\pm26.6$             \\
        17327       & lf4p02010 & COS        & G130M             & 60645.253                        & \phantom{0}0.481  & $-4.9\pm0.7$                         & $130.7\pm25.7$                      & $109.8\pm40.9$                       \\
        17327       & lf4p02040 & COS        & G130M             & 60645.292                        & \phantom{0}0.481  & $27.8\pm0.9$                         & $142.6\pm20.7$                      & $103.5\pm36.8$                       \\
        17625       & lfc901070 & COS        & G130M             & 60867.232                        & -0.182            & $-4.1\pm0.6$                         & $183.0\pm15.3$                      & \phantom{0}$10.7\pm24.9$             \\
        17625       & lfc9010a0 & COS        & G130M             & 60867.279                        & -0.181            & $-1.1\pm0.8$                         & $198.6\pm19.2$                      & \phantom{0}$27.4\pm27.4$             \\
        18177       & lfqw04020 & COS        & G130M             & 60968.168                        & 0.423             & $-2.0\pm0.8$                         & \phantom{0}$95.6\pm25.3$            & $138.0\pm8.5$                        \\
        18177       & lfqw04020 & COS        & G130M             & 60968.217                        & 0.424             & $-5.5\pm0.8$                         & \phantom{0}$101.5\pm31.7$           & $134.6\pm10.3$                       \\
        \hline \\
        \end{tabular}
        \begin{minipage}{0.95\linewidth}
            \ignorespaces 
            Notes: $^\dagger$ Velocity shift needed to align the \ion{C}{II} 1335 ISM lines. The calculated shift is relative to the STIS spectrum with ID o4wr11020.
        \end{minipage}
        
    \end{table}

    \begin{longtable}{lcclccccc}
 	          \caption{List of spectra from HST UV observations of AzV 75 containing the \ion{C}{IV} resonance line.}\label{tab:CIV_RVs} \\
 	        \hline \hline \rule{0cm}{2.2ex}%
            Proposal    & Spec. ID  & Instrument & Grating           & \multicolumn{1}{c}{Start Time}   & Phase             & $\varv_\mathrm{corr}^\dagger$        & RV$_1^\ddagger$                             & RV$_2^\ddagger$                              \\
                    &           &            &                   & \multicolumn{1}{c}{[JD-2400000.5]}&                  & \multicolumn{1}{c}{[km\,s$^{-1}$]}  & [km\,s$^{-1}$]                      & [km\,s$^{-1}$]                       \\
            \hline\rule{0cm}{2.2ex}%
            \endfirsthead 
            
            \multicolumn{9}{l}{{{\normalsize \tablename\ \thetable{} continued.}}} \vspace{0.3cm}\\
 	        \hline \hline \rule{0cm}{2.2ex}%
            Proposal    & Spec. ID  & Instrument & Grating           & \multicolumn{1}{c}{Start Time}   & Phase             & $\varv_\mathrm{corr}^\dagger$        & RV$_1^\ddagger$                              & RV$_2^\ddagger$                               \\
                    &           &            &                   & \multicolumn{1}{c}{[JD-2400000.5]}&                  & \multicolumn{1}{c}{[km\,s$^{-1}$]}  & [km\,s$^{-1}$]                      & [km\,s$^{-1}$]                       \\
            \hline\rule{0cm}{2.2ex}%
 			\endhead
    
     		\hline
     	    \multicolumn{9}{l}{{{
            \begin{minipage}{0.95\linewidth}
                \ignorespaces 
                \rule{0cm}{2.8ex} Notes: $^\dagger$ Velocity shift needed to align the \ion{Si}{II}$\lambda\lambda1526.72,1533.45$ ISM lines. The calculated shift is relative to the STIS spectrum with ID o4wr11020. $^\ddagger$ Radial velocity shift of the primary and secondary according to the orbital solution obtained with the PHOEBE code.
            \end{minipage}}}}\\
            \endfoot
                
     		\hline
     	    \multicolumn{9}{l}{{{
            \begin{minipage}{0.95\linewidth}
                \ignorespaces 
                \rule{0cm}{2.8ex} Notes: $^\dagger$ Velocity shift needed to align the \ion{Si}{II}$\lambda\lambda1526.72,1533.45$ ISM lines. The calculated shift is relative to the STIS spectrum with ID o4wr11020. $^\ddagger$ Radial velocity shift of the primary and secondary according to the orbital solution obtained with the PHOEBE code.
            \end{minipage}}}}\\
            \endlastfoot
            
        7437 & o4wr11020 & STIS & E140M & 51322.323 & \phantom{0}0.307 & \phantom{0}\phantom{0}$0.0\pm0.1$ & \phantom{0}76.4 & 177.1\\
        12805 & lbxk51020 & COS & G160M & 56181.908 & -0.412 & \phantom{0}$-3.3\pm0.2$ & 142.3 & \phantom{0}91.2\\
        13122 & lc6a01030 & COS & G160M & 56367.026 & -0.297 & \phantom{0}$-2.3\pm0.2$ & 167.4 & \phantom{0}58.5\\
        13122 & lc6a01040 & COS & G160M & 56367.081 & -0.297 & \phantom{0}\phantom{0}$0.9\pm0.2$ & 167.5 & \phantom{0}58.4\\
        13122 & lc6a01050 & COS & G140L & 56367.095 & -0.297 & $-36.4\pm0.9$ & 167.5 & \phantom{0}58.4\\
        13122 & lc6a01060 & COS & G140L & 56367.103 & -0.297 & $-34.8\pm1.6$ & 167.5 & \phantom{0}58.4\\
        13522 & lcgh01050 & COS & G160M & 56731.617 & -0.100 & \phantom{0}$-2.5\pm0.2$ & 202.2 & \phantom{0}13.2\\
        13522 & lcgh01060 & COS & G160M & 56731.656 & -0.100 & \phantom{0}\phantom{0}$1.3\pm0.2$ & 202.2 & \phantom{0}13.3\\
        13522 & lcgh01070 & COS & G140L & 56731.672 & -0.100 & $-58.7\pm1.0$ & 202.1 & \phantom{0}13.3\\
        13522 & lcgh01080 & COS & G140L & 56731.680 & -0.100 & $-24.7\pm1.1$ & 202.1 & \phantom{0}13.3\\
        13635 & lcil3a020 & COS & G140L & 56853.646 & -0.365 & \phantom{0}$47.7\pm1.7$ & 152.8 & \phantom{0}77.6\\
        13635 & lcil3a030 & COS & G140L & 56853.651 & -0.365 & \phantom{0}$48.5\pm1.5$ & 152.8 & \phantom{0}77.6\\
        13635 & lcil3a040 & COS & G140L & 56853.655 & -0.365 & \phantom{0}$54.5\pm1.1$ & 152.8 & \phantom{0}77.6\\
        13635 & lcil3a050 & COS & G140L & 56853.705 & -0.365 & \phantom{0}$33.2\pm1.3$ & 152.9 & \phantom{0}77.5\\
        13635 & lcil3a060 & COS & G140L & 56853.710 & -0.365 & \phantom{0}$39.2\pm1.1$ & 152.9 & \phantom{0}77.5\\
        13635 & lcil3a070 & COS & G140L & 56853.714 & -0.365 & \phantom{0}$36.1\pm0.9$ & 152.9 & \phantom{0}77.5\\
        13635 & lcil3a080 & COS & G140L & 56853.719 & -0.364 & \phantom{0}$48.0\pm1.3$ & 152.9 & \phantom{0}77.5\\
        13635 & lcil3a090 & COS & G140L & 56853.723 & -0.364 & \phantom{0}$35.0\pm0.9$ & 152.9 & \phantom{0}77.5\\
        13635 & lcil3a0a0 & COS & G140L & 56853.772 & -0.364 & \phantom{0}$56.5\pm1.9$ & 152.9 & \phantom{0}77.4\\
        13635 & lcil3a0b0 & COS & G140L & 56853.776 & -0.364 & \phantom{0}$39.1\pm1.6$ & 152.9 & \phantom{0}77.4\\
        13635 & lcil3a0c0 & COS & G140L & 56853.781 & -0.364 & $-32.6\pm1.8$ & 152.9 & \phantom{0}77.4\\
        13635 & lcil3a0d0 & COS & G140L & 56853.785 & -0.364 & \phantom{0}$37.4\pm1.0$ & 153.0 & \phantom{0}77.4\\
        13931 & lcrs01040 & COS & G160M & 57093.269 & \phantom{0}0.079 & \phantom{0}$17.6\pm0.2$ & \phantom{0}29.6 & 238.2\\
        13931 & lcrs01050 & COS & G160M & 57093.331 & \phantom{0}0.079 & \phantom{0}$21.2\pm0.2$ & \phantom{0}29.5 & 238.3\\
        13931 & lcrs51040 & COS & G160M & 57216.570 & -0.178 & \phantom{0}$-6.3\pm0.2$ & 193.3 & \phantom{0}24.8\\
        13931 & lcrs51050 & COS & G160M & 57216.631 & -0.178 & \phantom{0}$-3.9\pm0.2$ & 193.4 & \phantom{0}24.7\\
        13969 & lcpt01020 & COS & G140L & 57221.218 & -0.150 & $-15.3\pm1.3$ & 198.3 & \phantom{0}18.3\\
        13969 & lcpt01030 & COS & G140L & 57221.246 & -0.150 & \phantom{0}$12.5\pm1.2$ & 198.3 & \phantom{0}18.3\\
        14437 & ld2g51050 & COS & G160M & 57562.035 & -0.097 & \phantom{0} $-7.4\pm0.2$ & 201.9 & \phantom{0}13.6\\
        14437 & ld2g51060 & COS & G160M & 57562.075 & -0.096 & \phantom{0}$-2.7\pm0.2$ & 201.9 & \phantom{0}13.6\\
        14437 & ld2g51070 & COS & G140L & 57562.092 & -0.096 & $-58.5\pm1.1$ & 201.9 & \phantom{0}13.6\\
        14437 & ld2g51080 & COS & G140L & 57562.099 & -0.096 & $-10.9\pm1.1$ & 201.9 & \phantom{0}13.6\\
        14855 & ldc301050 & COS & G160M & 57830.535 & -0.479 & \phantom{0}$-4.7\pm0.2$ & 128.6 & 109.2\\
        14855 & ldc301060 & COS & G160M & 57830.548 & -0.479 & \phantom{0}\phantom{0}$0.2\pm0.2$ & 128.6 & 109.1\\
        14855 & ldc301070 & COS & G140L & 57830.564 & -0.479 & $-40.7\pm0.9$ & 128.6 & 109.1\\
        14855 & ldc301080 & COS & G140L & 57830.572 & -0.479 & \phantom{0}\phantom{0}$7.8\pm0.8$ & 128.6 & 109.1\\
        15366 & ldm701020 & COS & G160M & 57975.289 & \phantom{0}0.393 & \phantom{0}$-4.2\pm0.2$ & 100.1 & 146.2\\
        15366 & ldm701030 & COS & G160M & 57975.357 & \phantom{0}0.394 & \phantom{0}\phantom{0}$0.1\pm0.2$ & 100.2 & 146.1\\
        15385 & ldq701020 & COS & G160M & 58189.146 & -0.318 & \phantom{0}$-1.8\pm0.2$ & 163.4 & \phantom{0}63.8\\
        15385 & ldq701030 & COS & G160M & 58189.187 & -0.318 & \phantom{0}\phantom{0}$2.1\pm0.2$ & 163.4 & \phantom{0}63.7\\
        15385 & ldq701060 & COS & G140L & 58189.267 & -0.317 & $-22.6\pm1.0$ & 163.6 & \phantom{0}63.6\\
        15385 & ldq701070 & COS & G140L & 58189.277 & -0.317 & \phantom{0}$55.3\pm1.1$ & 163.6 & \phantom{0}63.6\\
        15536 & lduh01020 & COS & G160M & 58561.210 & -0.076 & \phantom{0}$-5.7\pm0.3$ & 198.0 & \phantom{0}18.6\\
        15536 & lduh01060 & COS & G140L & 58561.325 & -0.076 & $-57.8\pm1.5$ & 197.8 & \phantom{0}18.9\\
        15536 & lduh01070 & COS & G140L & 58561.335 & -0.076 & \phantom{0}$22.5\pm1.4$ & 197.8 & \phantom{0}18.9\\
        15536 & lduh51020 & COS & G160M & 58659.717 & -0.483 & \phantom{0}$-5.8\pm0.2$ & 128.0 & 109.9\\
        15536 & lduh51030 & COS & G160M & 58659.772 & -0.482 & \phantom{0}$-3.4\pm0.2$ & 128.1 & 109.8\\
        15536 & lduh51060 & COS & G140L & 58659.850 & -0.482 & $-40.5\pm0.9$ & 128.2 & 109.7\\
        15536 & lduh51070 & COS & G140L & 58659.860 & -0.482 & \phantom{0}$42.4\pm1.8$ & 128.2 & 109.7\\
        15774 & le0r01020 & COS & G160M & 58928.401 & \phantom{0}0.136 & \phantom{0}$-3.3\pm0.2$ & \phantom{0}32.5 & 234.4\\
        15774 & le0r01030 & COS & G160M & 58928.447 & \phantom{0}0.136 & \phantom{0}$-1.7\pm0.2$ & \phantom{0}32.6 & 234.3\\
        15774 & le0r01060 & COS & G140L & 58928.513 & \phantom{0}0.137 & $-28.2\pm0.9$ & \phantom{0}32.7 & 234.2\\
        15774 & le0r01070 & COS & G140L & 58928.523 & \phantom{0}0.137 & \phantom{0}$20.6\pm1.4$ & \phantom{0}32.7 & 234.2\\
        16467 & leij01020 & COS & G140L & 59288.393 & \phantom{0}0.305 & $-53.5\pm1.0$ & \phantom{0}79.0 & 173.7\\
        16325 & lefc01020 & COS & G160M & 59414.730 & \phantom{0}0.066 & \phantom{0}$-7.7\pm0.2$ & \phantom{0}32.8 & 234.1\\
        16325 & lefc01030 & COS & G160M & 59414.785 & \phantom{0}0.067 & \phantom{0}$-2.7\pm0.2$ & \phantom{0}32.6 & 234.3\\
        16325 & lefc01060 & COS & G140L & 59414.863 & \phantom{0}0.067 & $-51.4\pm2.1$ & \phantom{0}32.4 & 234.5\\
        16325 & lefc01070 & COS & G140L & 59414.873 & \phantom{0}0.067 & \phantom{0}$24.9\pm1.3$ & \phantom{0}32.4 & 234.6\\
        16907 & letc01010 & COS & G160M & 59648.011 & \phantom{0}0.472 & \phantom{0}\phantom{0}$0.0\pm1.1$ & 118.5 & 122.3\\
        16907 & letc01020 & COS & G160M & 59648.090 & \phantom{0}0.473 & \phantom{0}\phantom{0}$0.0\pm0.2$ & 118.6 & 122.2\\
        16534 & leko01020 & COS & G160M & 59760.624 & \phantom{0}0.151 & \phantom{0}$-8.8\pm0.2$ & \phantom{0}36.3 & 229.4\\
        16534 & leko01030 & COS & G160M & 59760.669 & \phantom{0}0.151 & \phantom{0}$-5.1\pm0.2$ & \phantom{0}36.4 & 229.3\\
        16534 & leko01060 & COS & G140L & 59760.739 & \phantom{0}0.151 & $-23.2\pm2.0$ & \phantom{0}36.5 & 229.2\\
        16534 & leko01070 & COS & G140L & 59760.749 & \phantom{0}0.151 & \phantom{0}$-8.5\pm1.0$ & \phantom{0}36.5 & 229.1\\
        17250 & lf2301020 & COS & G160M & 60140.588 & \phantom{0}0.440 & \phantom{0}$-3.9\pm0.2$ & 111.5 & 131.4\\
        17250 & lf2301030 & COS & G160M & 60140.605 & \phantom{0}0.440 & $-10.9\pm0.2$ & 111.6 & 131.3\\
        17250 & lf2301040 & COS & G160M & 60140.650 & \phantom{0}0.440 & \phantom{0}$-1.5\pm0.2$ & 111.6 & 131.3\\
        17250 & lf2301050 & COS & G160M & 60140.668 & \phantom{0}0.440 & \phantom{0}$-7.6\pm0.2$ & 111.6 & 131.2\\
        17250 & lf2301080 & COS & G140L & 60140.739 & \phantom{0}0.441 & $-65.6\pm1.3$ & 111.7 & 131.1\\
        17250 & lf2301090 & COS & G140L & 60140.774 & \phantom{0}0.441 & $-35.6\pm1.3$ & 111.8 & 131.0\\
        17327 & lf4p01040 & COS & G160M & 60499.207 & -0.399 & \phantom{0}$-1.3\pm0.2$ & 146.6 & \phantom{0}85.7\\
        17327 & lf4p02020 & COS & G140L & 60645.278 & \phantom{0}0.481 & \phantom{0}$-7.7\pm1.3$ & 120.8 & 119.4\\
        17327 & lf4p02030 & COS & G140L & 60645.288 & \phantom{0}0.481 & $134.6\pm11$ & 120.8 & 119.3\\
        17625 & lfc901020 & COS & G160M & 60867.113 & -0.182 & \phantom{0}$-5.9\pm0.2$ & 193.5 & \phantom{0}24.5\\
        17625 & lfc901030 & COS & G160M & 60867.148 & -0.182 & $-12.4\pm0.2$ & 193.6 & \phantom{0}24.4\\
        17625 & lfc901040 & COS & G160M & 60867.162 & -0.182 & \phantom{0}$-2.9\pm0.3$ & 193.6 & \phantom{0}24.4\\
        17625 & lfc901050 & COS & G160M & 60867.180 & -0.182 & \phantom{0}$-8.3\pm0.2$ & 193.6 & \phantom{0}24.4\\
        17625 & lfc901080 & COS & G140L & 60867.244 & -0.182 & $-73.9\pm1.9$ & 193.7 & \phantom{0}24.3\\
        17625 & lfc901090 & COS & G140L & 60867.254 & -0.182 & $-43.9\pm1.1$ & 193.7 & \phantom{0}24.3\\
    \end{longtable}
     
\section{Additional Figures}
   
    \begin{figure*}[hb]
        \centering
        \begin{minipage}[b]{.45\textwidth}
        \begin{tikzpicture}[scale=1]
            \node[anchor=south] at (0, 0) {\includegraphics[trim= 6.5cm 0cm 2.5cm -0.15cm ,clip, width=\textwidth]{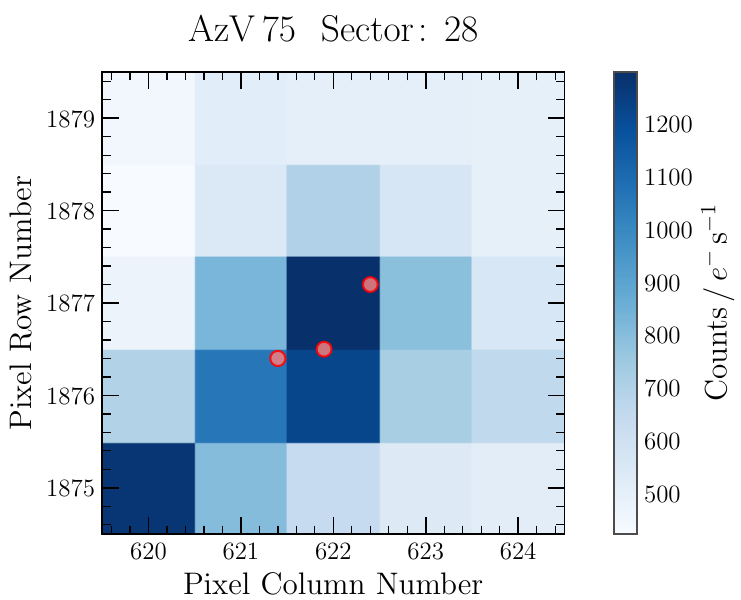}};
            \node at (0, 4.55) {\fontsize{10}{16}\color{red}\selectfont AzV\,75};
            \draw[-latex,red,thick] (0,4.3) -- (0.,3.9);
            \node at (-1., 1.6) {\fontsize{10}{16}\color{red}\selectfont OGLE SMC-SC4 19933};
            \draw[-latex,red,thick] (-1.3,1.8) -- (-1.05,2.7);
            \node at (0.2, 2.2) {\fontsize{10}{16}\color{red}\selectfont UCAC2 1078504};
            \draw[-latex,red,thick] (-0.2,2.4) -- (-0.48,2.8);
        \end{tikzpicture}    
        \caption{The TESS Sector 28 field of view. Red circles mark sources brighter than $G<\SI{14}{mag}$. The central pixel, containing AzV 75, includes one other source of comparable brightness (at lower edge of pixel). Colours represent average counts received in each pixel.}
        \label{fig:sector28}
        \end{minipage}\hspace{0.2cm}
        \begin{minipage}[b]{.45\textwidth}
        \begin{tikzpicture}[scale=1]
            \node[anchor=south] at (0, 0) {\includegraphics[trim= 6.5cm 0cm 2.5cm -0.15cm ,clip, width=\textwidth]{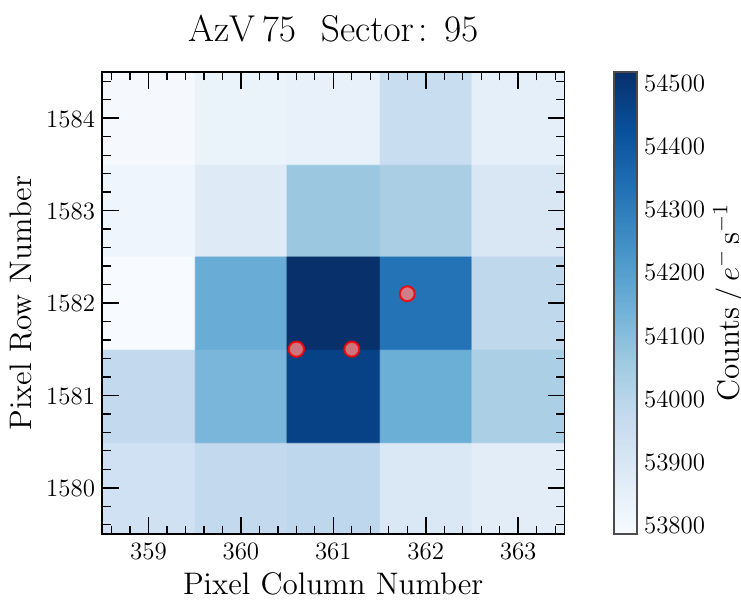}};
            \node at (0+0.4, 4.55) {\fontsize{10}{16}\color{red}\selectfont AzV\,75};
            \draw[-latex,red,thick] (0+0.4,4.3) -- (0.+0.4,3.9);
            \node at (-1.+0.15, 1.6) {\fontsize{10}{16}\color{red}\selectfont OGLE SMC-SC4 19933};
            \draw[-latex,red,thick] (-1.3+0.15,1.8) -- (-1.05+0.15,2.7);
            \node at (0.2+0.4, 2.2) {\fontsize{10}{16}\color{red}\selectfont UCAC2 1078504};
            \draw[-latex,red,thick] (-0.2+0.4,2.4) -- (-0.48+0.4,2.8);
        \end{tikzpicture}    
        \caption{The TESS Sector 95 field of view. Red circles mark sources brighter than $G<\SI{14}{mag}$. The pixel, containing AzV 75, includes no other source of comparable brightness. Colours represent average counts received in each pixel. Note that the flux count in this sector is about a factor 50 higher, due to scattered light of the earth and moon \citep{TESS_DR_sector95}.}
        \label{fig:sector95}
        \end{minipage}
    \end{figure*}

\clearpage

   \begin{figure*}
   \centering
    \includegraphics[trim= 3.4cm 3.5cm 3.6cm 4.4cm ,clip, width=\textwidth]{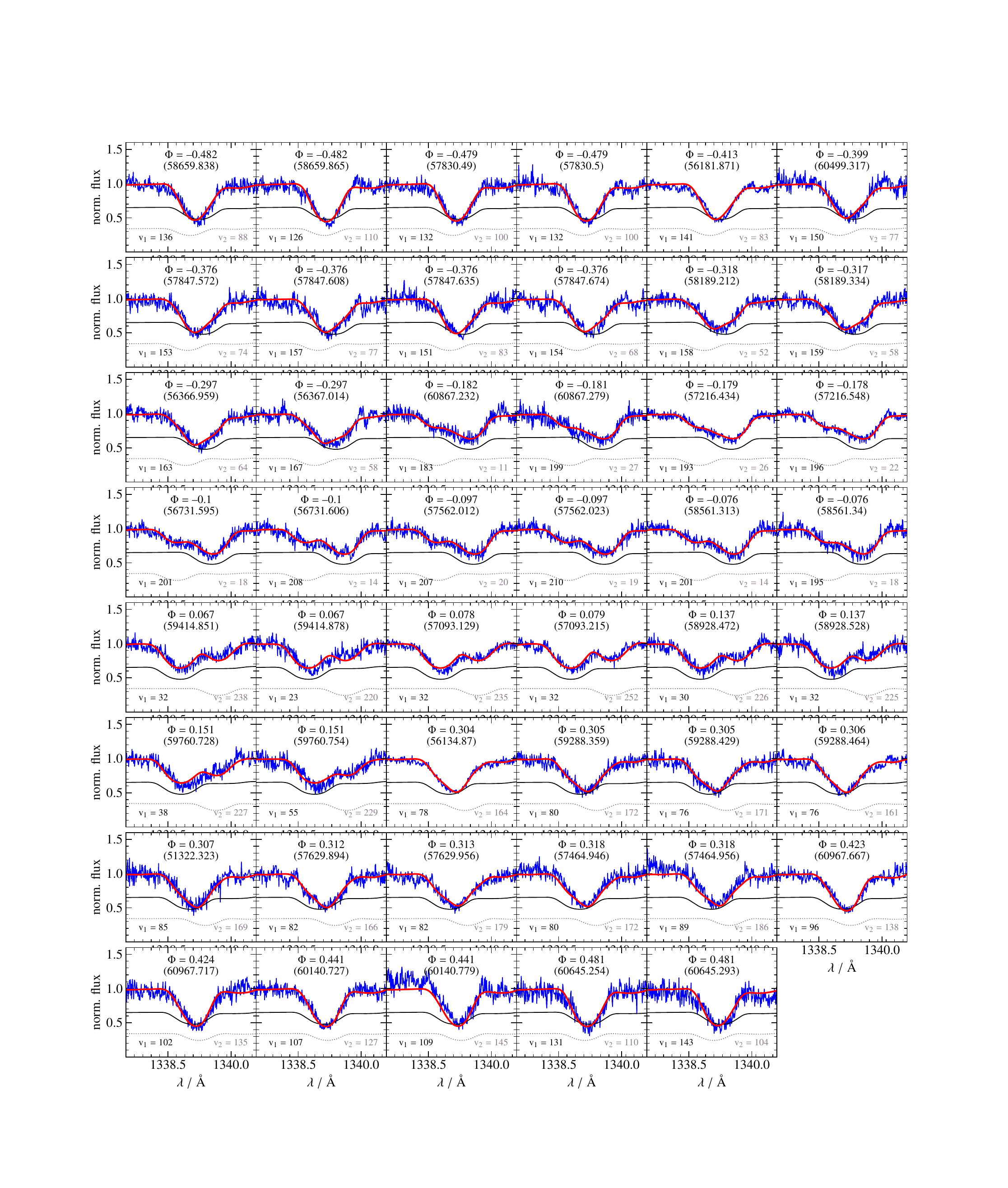}
      \caption{Two stellar model spectra (solid black and dashed purple lines, combined result shown as red solid line) fitted to the \ion{O}{IV}\,$\lambda1338.61$ photospheric line in each observed spectrum (blue lines) from the HST UV data set. The synthetic stellar spectra use approximations to true stellar parameters, but are sufficient to obtain well-fitted radial velocities for each component at various orbital phases.}
         \label{fig:phases}
   \end{figure*}
   
   \begin{figure*}
   \centering
    \includegraphics[trim= 0.6cm 0.5cm 0.9cm 0.cm ,clip, width=0.8\linewidth]{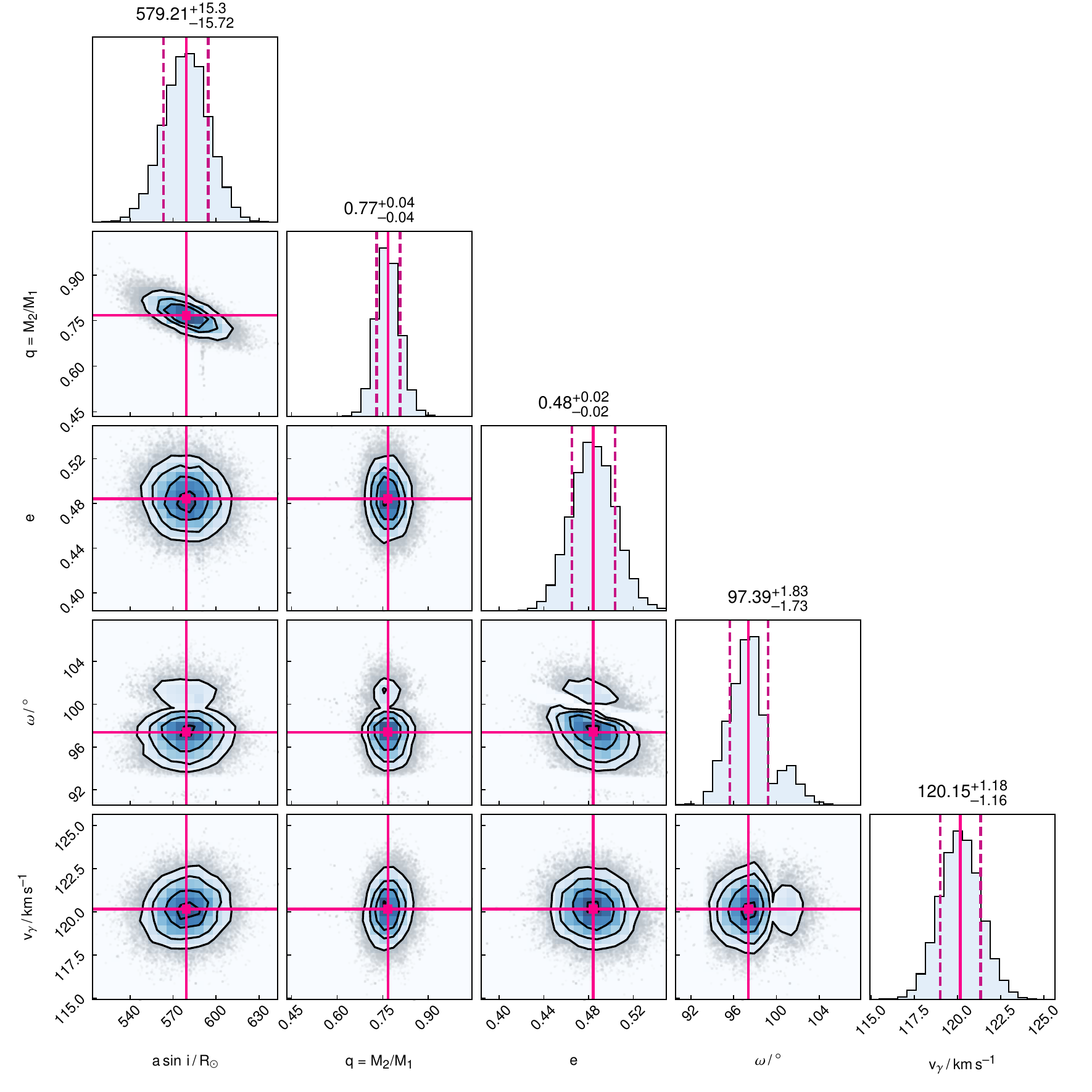}
      \caption{Corner plot showing the posterior distributions of orbital parameters derived from HST UV RV measurements for each component of the AzV 75 system using PHOEBE's MCMC sampler. The reported values are the mean of the posterior distributions.}
         \label{fig:RVposteriors}
   \end{figure*}
   
   \twocolumn
   \begin{figure*}
   \centering
    \includegraphics[trim= 2cm 23.4cm 2.5cm 4cm ,clip, width=0.9\linewidth]{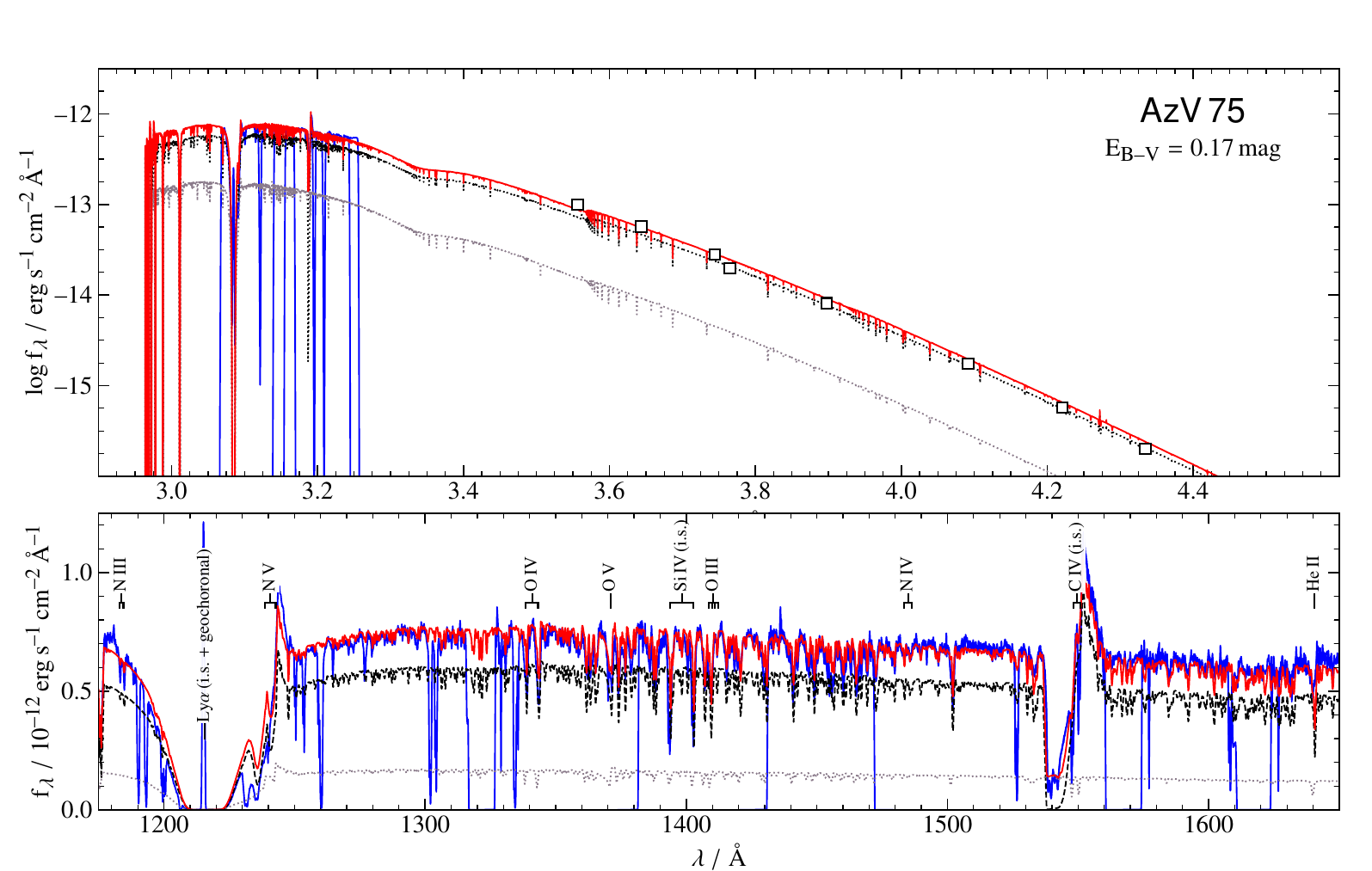}
      \caption{Comparison of the selected PoWR stellar atmosphere grid models to the observations of AzV~75. \textit{Top panel:} Spectral energy distribution. Open squares mark the U, B, V, and I photometry from \citet{Bonanos2010}, Gaia G magnitude from \citet{Gaia2020}, and J, H, and K from \citet{Cutri2003}. Blue lines mark the flux-calibrated spectra. The black dashed line is the synthetic model of the primary, and the purple dotted line the synthetic model of the secondary. The red line is the combined synthetic spectrum. The synthetic spectra are corrected for reddening from a Galactic foreground modeled with the law of \citet{Seaton1979} and $E_\mathrm{B-V}=0.06\,\mathrm{mag}$ and the background of the SMC using the law of \citet{Gordon2003} with $E_\mathrm{B-V}=0.11$. \textit{Bottom panel:} Flux-calibrated HST UV spectrum taken at 2456732.1\,HJD (blue), compared to the combined (red) and individual (black and purple) synthetic spectra. Note that these are not tailored models but from a recalculated model grid.}
         \label{fig:PoWR}
   \end{figure*}

    \begin{figure*}
        \centering
        \begin{minipage}[b]{.45\textwidth}
        \includegraphics[trim= 0.6cm 0.5cm 0.9cm 0.cm ,clip, width=\linewidth]{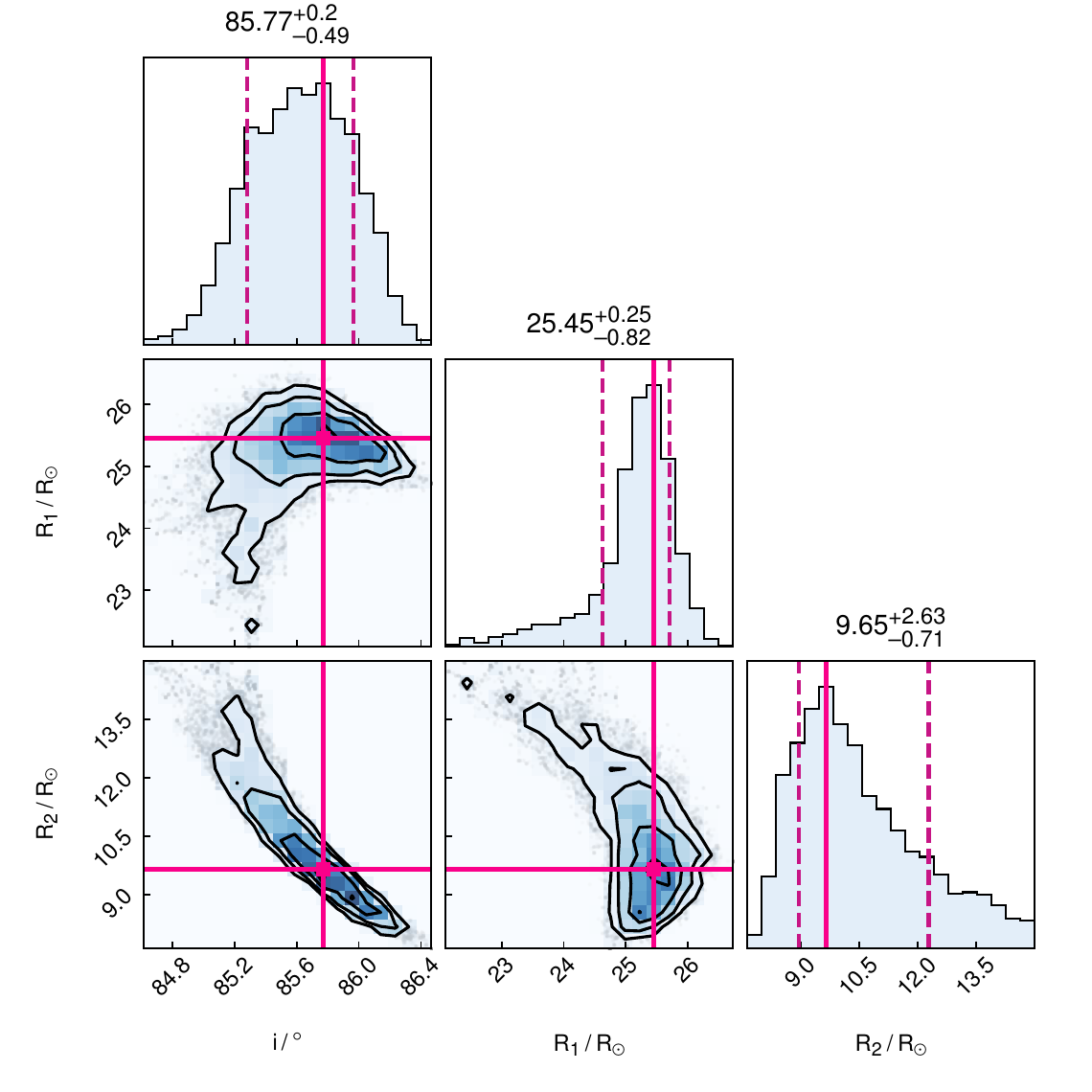}
        \caption{Corner plot showing the posterior distributions of orbital parameters derived from the ASAS-SN g-Band and TESS sector 95 light curves using PHOEBE's MCMC sampler. Since the distributions do not have a symmetric shape due to their dependence on each other, the reported values are the modes of the posterior distribution instead of the mean.}
        \label{fig:LCposteriors}
        \end{minipage}\hspace{0.2cm}
        \begin{minipage}[b]{.45\textwidth}
        \includegraphics[trim= 0.6cm 0.5cm 0.9cm 0.cm ,clip, width=0.8\linewidth]{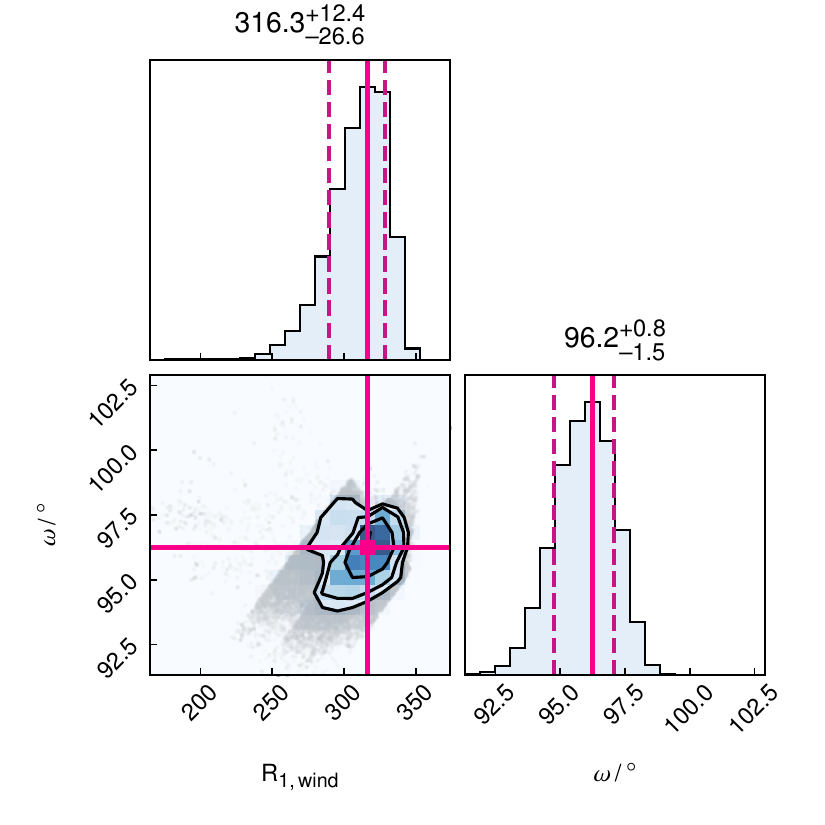}
        \caption{Corner plot showing the posterior distributions of the radial extent of the \ion{C}{IV} resonance line formation region derived from the HST spectra. The reported values are the modes of the posterior distribution.}
        \label{fig:CIVposterior}
        \end{minipage}
    \end{figure*}
   
   \begin{figure*}
   \centering
    \includegraphics[trim= 7cm 0.0cm 8cm 0.cm ,clip, width=\linewidth]{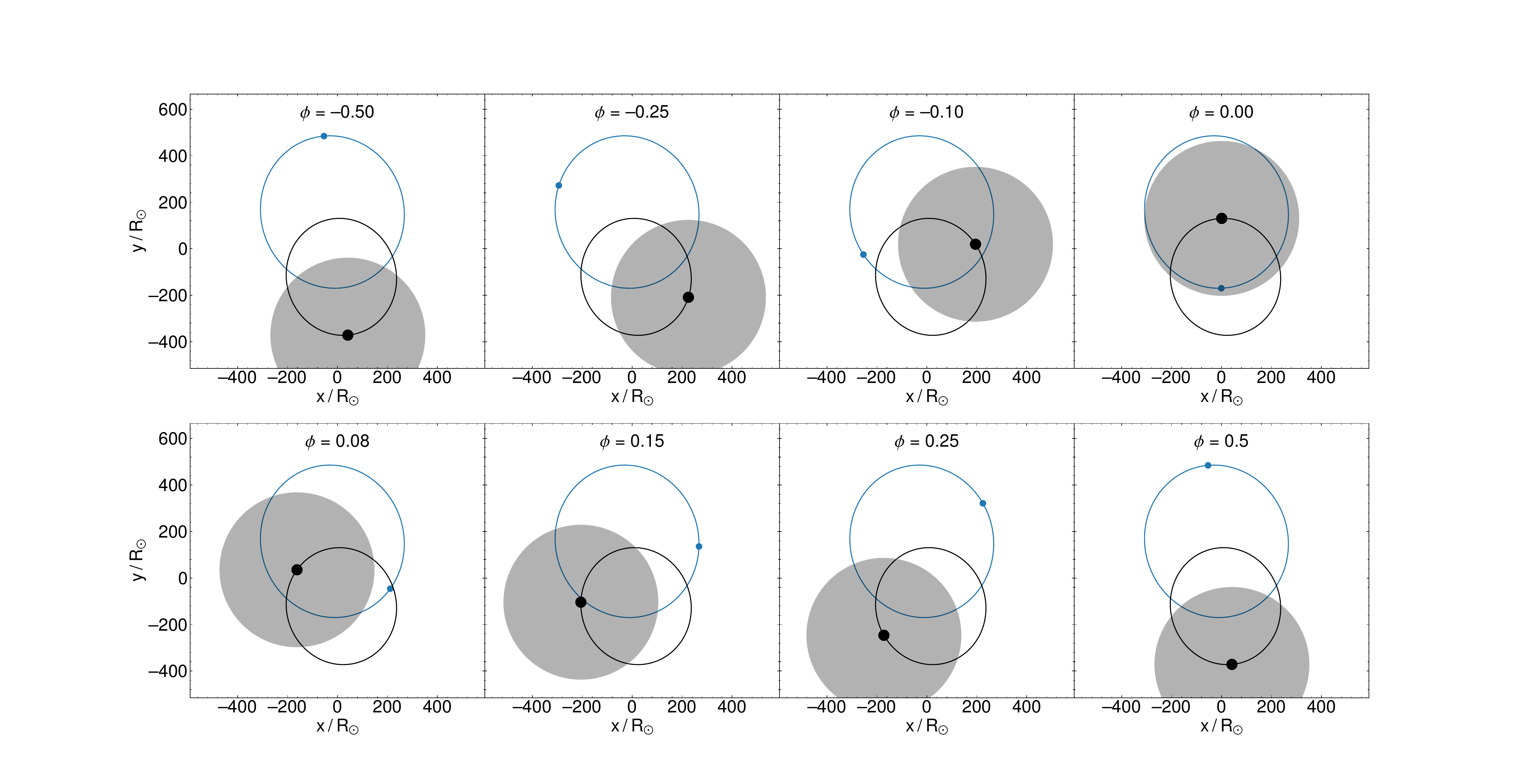}
      \caption{Illustration of the orbital configuration of the system in the orbital plane at several phases. The black and blue curves indicate the trajectories of the primary and secondary stars, respectively. The current position of the primary is marked by the black dot, and the grey shaded region denotes the \ion{C}{IV} line‑formation zone in the primary’s wind. The blue dot marks the location of the secondary. The secondary star's putative wind line-formation zone has been omitted for clarity. The observer is situated at $x=0$, at the bottom of the diagram.}
         \label{fig:orbit}
   \end{figure*}
   
   \begin{figure*}
   \centering
    \includegraphics[trim= 1.1cm 0.8cm 2.7cm 1.9cm ,clip,width=\linewidth]{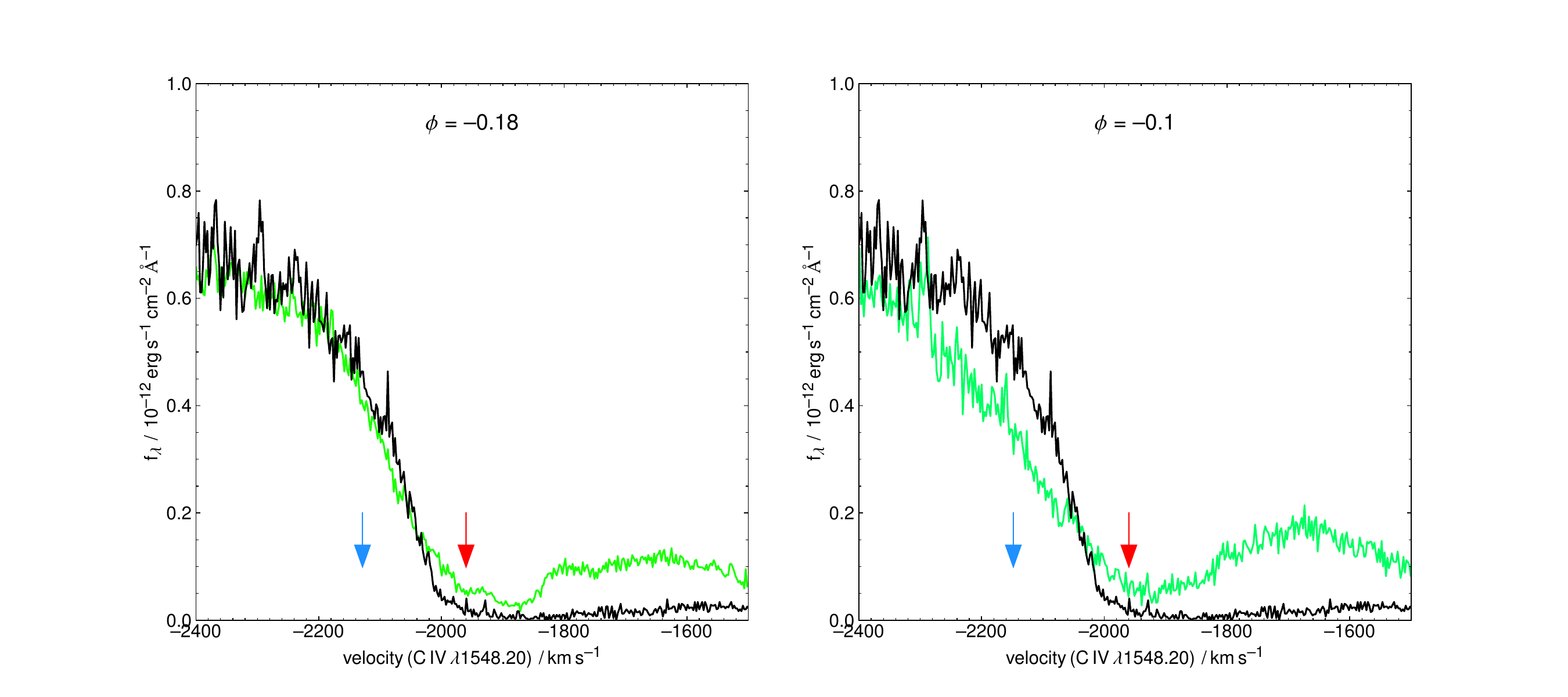}
      \caption{Zoom-in on the blue edge of the C IV absorption trough in velocity space with respect to \ion{C}{IV}\,$\lambda1548.20$ in the rest frame of the primary star. A reference spectrum taken at phase $\phi=-0.48$ when the primary star’s wind occults the secondary is shown in black. A spectrum taken at the indicated phase is colour coded according to the colour bar presented in Fig.~\ref{fig:vinf_edge}. A red arrow marks the terminal wind velocity of the primary, while a blue arrow marks the relative RV shift of the secondary with respect to the primary.}
         \label{fig:vinf_edge_selected}
   \end{figure*}

\end{appendix}

\end{document}